\documentclass[12pt]{article}
\usepackage{amsmath}
\usepackage{amssymb}
\begin{document}
\begin{center}
{\bf {\large{Modified 3D Massive Abelian 2-From Theory with a Single Pseudo-Scalar Field as a Phantom Field: BRST Approach }}} 

\vskip 2cm

{\sf  S. K. Panja$^{(a)}$,  E. Harikumar$^{(a)}$, R. P. Malik$^{(b,c)}$}\\

\vskip 0.1cm

$^{(a)}$ School of Physics, University of Hyderabad,\\
Central University P. O., Gachibowli, Hyderabad-500 046, Telangana, India\\

\vskip 0.1cm

$^{(b)}$ {\it Physics Department, Institute of Science,}\\
{\it Banaras Hindu University (BHU), Varanasi-221 005, Uttar Pradesh (U.P.), India}\\

\vskip 0.1cm

$^{(c)}$ {\it DST Centre for Interdisciplinary Mathematical Sciences,}\\
{\it Institute of Science, Banaras Hindu University, Varanasi-221 005, U. P., India}\\
{\small {\sf {e-mails: sumanpanja19@gmail.com; eharikumar@uohyd.ac.in; rpmalik1995@gmail.com}}}
\end{center}

\vskip 1.5 cm

\noindent
{\bf Abstract:}
We obtain the off-shell nilpotent Becchi-Rouet-Stora-Tyutin (BRST) and anti-BRST symmetry transformations (corresponding to the infinitesimal classical gauge symmetry transformations) for the {\it modified} massive three $(2+1)$-dimensional (3D) Abelian 2-form gauge theory with a single pseudo-scalar field. 
The {\it latter} field (having the negative kinetic term and a well-defined rest mass) has already been shown (i) to exist in the 
{\it modified} version of the standard 3D St${\ddot u}$ckelberg formalism (on the solid mathematical grounds),  (ii) to be a possible candidate
for the ``phantom" field of some of the cosmological models of the Universe, and 
(iii) to be a possible candidate for dark matter. These {\it three} results have been indicated in our earlier work.
A couple of novel observations in our present endeavor are (i) the observation that, even though the pseudo-scalar
field does {\it not} transform under the gauge and (anti-)BRST symmetry transformations, it appears in the first-class constraints which annihilate the
physical states at the quantum level, and (ii) the {\it Noether} conserved (anti-)BRST charges are found to be {\it non-nilpotent}. In our present investigation,
the key results are the derivations of (i) the coupled (but equivalent) BRST and anti-BRST invariant Lagrangian densities, (ii) the conserved 
and off-shell nilpotent 
versions of the (anti-)BRST charges and the conserved ghost charge, (iii) the (anti-)BRST invariant Curci-Ferrari (CF) type restrictions,  
(iv) the standard BRST algebra amongst the conserved and nilpotent (anti-)BRST charges and conserved ghost charge, and (v) the explicit
BRST-quantization of our 3D field-theoretic system.



%1
\newpage
\section{Introduction}
The modern developments in the research activities, connected with the ideas behind (super)string theories (see, e.g. [1-5] for details), have brought together a set of very active researchers (i) in the domain of theoretical high energy physics (THEP), and (ii) in the realm of pure mathematics, on an intellectual platform where both sets of researchers have benefited from each-other. As far as the research activities in the realm of quantum field theories are concerned, mention can be made of the theoretical works done in the areas of (i) the topological field theories (see, e.g. [6-9] and references therein), (ii) the higher spin gauge theories (see, e.g. [10,11] and references therein), (iii) the supersymmetric Yang-Mills theories (see, e.g. [12,13] and references therein), etc., where there have been convergence of ideas from THEP and pure mathematics. We have devoted time, during the last few years, on the study of the higher $p$-form $(p=2,3 \ldots)$ gauge theories\footnote{The most successful theory in THEP is the standard model of particle physics (SMPP) which is based on the interacting non-Abelian 1-form 
(i.e. $p=1$) gauge theory where there has been stunning degree of agreements between theory and experiment. However, it is plagued with a large number of adjustable parameters and it does not include the theory of gravity (as far as its theoretical reach and range for the unification scheme is concerned). Moreover, it has been experimentally verified that the weakly interacting neutrinos have masses. This observation nullifies one of
the basic principles on which the SMPP is based.}
which is inspired by the ideas behind (super)string theories because the higher $p$-form $(p=2,3, \ldots)$ basic fields appear in the quantum excitations of the (super)strings. We have been able to establish that the St${\ddot u}$ckelberg-modified massive and massless Abelian $p$-form $(p=1,2,3)$ gauge theories, in $D=2 p$ dimensions of spacetime, are the field-theoretic examples for Hodge theory (see, e.g. [14-16] and references therein) within the framework of Becchi-Rouet-Stora-Tyutin (BRST) formalism [17-20] where the discrete and continuous symmetries (and corresponding Noether conserved charges) have been able to provide the physical realizations of the de Rham cohomological operators (see. e.g. [21-24] for details) of differential geometry at 
the {\it algebraic} level. It is quite obvious, from the above discussions, that the field-theoretic models of Hodge theory are defined only in the 
{\it even} (i.e. $D=2 p$) dimensions of spacetime.

In our present investigation, we concentrate on the study of a specific massive higher $p$-form (i.e. $p=2$) gauge theory in the 
{\it odd} dimension (i.e. $D=3$) of spacetime. We have obtained a well-defined 3D Lagrangian density [cf. Eq. (1) below] for the modified massive Abelian 2-form gauge theory where we have exploited the modification [25] in the standard St${\ddot u}$ckelberg technique 
of replacement for the massive Abelian 2-form gauge field. In this modification, in addition to the presence of the standard St${\ddot u}$ckelberg vector field, we have a single pseudo-scalar field which is incorporated on the basis of solid mathematical arguments (see, e.g. [26] for details). The latter field appears in the theory with a negative kinetic term. However, it is endowed with a well-defined rest mass because it obeys the standard Klein-Gordon equation of motion\footnote{Such kinds of fields are ``exotic" fields and they have become quite popular in the realm of cosmological models of the Universe where they have been christened as the ``phantom" or ``ghost" fields. These fields (having the negative kinetic terms and well defined masses) are {\it also} possible candidates for dark matter.}. We have demonstrated the existence of the first-class constraints on {\it this} 
specific (i.e. St${\ddot u}$ckelberg-modified) theory. The latter generate the infinitesimal, local and continuous classical gauge symmetry transformations [cf. Eq. (4) below]. We have generalized these classical gauge symmetry transformations to their quantum counterparts as the BRST and anti-BRST symmetry transformations which are
respected by the coupled (but equivalent) Lagrangian densities. The {\it latter} owe their origin to the (anti-)BRST invariant Curci-Ferrari (CF) type restrictions (which are the hallmark of a properly BRST-quantized theory) and they also respect the ghost-scale symmetry transformations. We have derived, in our present endeavor, the conserved and off-shell nilpotent versions of the (anti-)BRST charges and the conserved ghost charge and demonstrated that they obey the 
{\it standard} BRST algebra. The (anti-)BRST invariant CF-type restrictions have been derived, in our present endeavor, from different theoretical angles. These derivations happen to be one of the highlights of our present investigation.

Against the backdrop of the above two paragraphs, it is crystal clear that our present investigation is 
{\it different} from the St${\ddot u}$ckelberg-modified massive Abelian 2-form (i.e. $p=2$) gauge theory which has been already established by us as a field-theoretic model for Hodge theory in $D=4$ dimensions of spacetime (see. e.g. $[14,25]$ for details) where (i) a pseudo-scalar field, and (ii) an axial-vector field have been shown to possess $(a)$ the negative kinetic terms, and $(b)$ the well-defined masses. In other words, we have been able to establish that there are two ``exotic" fields in the above modified massive 4D field-theoretic model for Hodge theory which play very decisive roles in the realm of the cyclic, bouncing and self accelerated cosmological models of the Universe (see, e.g. 
[27-32]) and they are also a set of possible candidates for dark matter (see, e.g. [33,34] for details). We firmly believe that all the modified massive $2 p$-dimensional Abelian $p$-form models for Hodge theory, within the framework of our BRST approach, are endowed with a tower of $p$-number of ``exotic" fields and it is not clear whether all of them are the most fundamental ``exotic" fields or only some of them. This is not the case with our present investigation where only 
a {{\it single} pseudo-scalar field has been shown to be an ``exotic" field. It is appropriate, at this juncture, to highlight point-by-point, some of the key differences between our present work and our earlier works on the 4D modified massive Abelian 2-form theory (see, e.g. $[14,25]$ ). First of all, we are dealing with an odd dimensional (i.e. $D=3$ ) modified massive Abelian 2-form theory which is not the case with our earlier works [14,25] in the 4D spacetime. Second, as pointed our earlier, there is only one ``exotic" field in our present theory unlike our previous works $[14,25]$ where there are two. Finally, in our present endeavor, there is no presence of an axial-vector field as an ``exotic" field as is the case in our earlier works.

Our present investigation is essential and important on the following counts. First of all, as pointed out earlier, we have shown the existence of a {\it single} pseudo-scalar field which appears in our theory with a negative kinetic term and possesses a well-defined rest mass. Hence, we have a single ``exotic" field in our theory. Second, the Noether 
conserved (anti-)BRST charges $Q_{(a) b}$ turn out to be non-nilpotent (i.e. $Q_{(a) b}^{2} \neq 0$). We have been able to obtain the nilpotent (i.e. $Q_{(A) B}^{2}=0$) versions of the (anti-)BRST charges $Q_{(A) B}$ which 
(i) participate in the standard BRST algebra (cf. Appendix A below for details), and (ii) lead to the derivations of the first-class constraints in their
operator form through the requirement of the physicality criteria w.r.t. them (cf. Appendix B below for details).
 Third, the key signature of a properly BRST-quantized theory is the existence of a set of (non-)trivial CF-type restrictions (see, e.g. $[35,36]$ and references therein). The {\it latter} for our theory are non-trivial and we have derived them from different theoretical angles. Finally, it appears to us that the pseudo-scalar field 
(having a negative kinetic term and a well-defined rest mass) is the most fundamental object as far as the existence of the ``exotic" fields that provide a set of possible candidates for the ``phantom'' field [27-32] as well as the dark matter
[33,34]. The massless limit of this field (with only a negative kinetic term) is one of the 
possible candidates for dark energy that has become quite popular in the realm of the cyclic, bouncing and self-accelerated cosmological models 
[27-32] which have been proposed to explain the accelerated expansion of the Universe.

At this crucial juncture, it is worthwhile to mention that, in the properly gauge-fixed Lagrangian densities [cf. Eqs. (23),(48)], we have the
(pseudo-)scalar fields $(\tilde \varphi)\phi$ in our theory {\it together}. However, a close look at the kinetic terms of these fields 
demonstrate that {\it both} of them have kinetic terms with the {\it opposite} signs.  Despite this discerning difference, both these fields obey 
the famous Klein-Gordon
EL-EoMs [cf. footnote associated with the gauge-fixed Lagrangian density (22)] implying that both these fields (i) carry equal well-defined mass, and (ii) are relativistic and elementary in nature. The field like the pseudo-scaler (PS) field  with (i) the negative kinetic term, and (ii) a well-defined rest 
mass, are very much sought after in the realm of the cyclic, bouncing and self-accelerated cosmological models of the Universe where such fields have been
called as the ``phantom'' field or ``ghost'' field (see, e.g. [27-32] for details). The presence pf such kinds of fields, ultimately,  leads to the 
existence of the negative pressure in a theory which is one of the  characteristic features of dark energy. The title, abstract and the main text of
the earlier works (see, e.g. [33,34] and references therein) indicate that the fields with negative kinetic terms (and well-defined masses or massless)  are
a set of possible candidate for dark matter/dark energy. In our present endeavor, the PS field is such a field that appears in the modification
of the standard St{\" u}ckelberg formalism for the 3D massive Abelian 2-form theory. This field, not only appears in the first-class constraints, it participates
in the (anti-)co-BRST symmetry transfigurations, too,  which have been shown in our earlier works [15,14,25]. However, such symmetries for our theory
[cf. Eq. (74) of our Sec. 6] are {\it yet} to be fully completed in our future endeavor(s). Hence, we firmly believe that 
the PS field of our theory is important from the point of view of
(i) the existence of the well-known phantom field(s), and  (ii) the possible candidate for dark matter/dark energy.

The theoretical materials of our present endeavor are organized as follows. In Sec. 2, we recapitulate the bare essentials of our earlier work on the modified 3D massive Abelian 2-form gauge theory where we (i) show the existence of a pseudo-scalar field in the 
{\it modification} of the standard St${\ddot u}$ckelberg formalism, (ii) derive the Noether conserved current and charge corresponding to the gauge symmetry transformations, and (iii) establish a connection between the conserved charge and the first-class constraints of the gauge theory. Our Sec. 3 is devoted to the discussion on the BRST symmetry transformations, perfectly 
BRST invariant Lagrangian density and the derivation of the BRST charge. The subject matter of our Sec. 4 is connected with the discussion on the anti-BRST symmetry transformations, 
perfectly anti-BRST invariant Lagrangian density and the derivation of the anti-BRST charge. Our Sec. 5 deals with the importance of the Curci-Ferrari (CF) type restrictions from different theoretical angles. Finally, we make some concluding remarks and point out the future scope and perspective of our present endeavor in our Sec. 6.

In our Appendix A, we derive the standard BRST algebra amongst the nilpotent versions of the (anti-)BRST charges and the conserved ghost charge. Our Appendix B deals with the physicality criterion w.r.t. the nilpotent BRST charge in a concise manner. The BRST-quantization of our present 3D field-theoretic system
is discussed in our Appendix C.

{\it Conventions and Notations}: The background 3D flat Minkowskian spacetime is endowed with the metric tensor $\eta_{\mu \nu}=\operatorname{diag}\,(+1,-1,-1)$ so that the dot product between two non-null vectors $U_{\mu}$ and $V_{\mu}$ is defined as: 
$U \cdot V=\eta_{\mu \nu} \,U^{\mu} \,V^{\nu} \equiv U_{0} V_{0}-U_{i} V_{i}$ where the Greek indices $\mu, \nu, \lambda \ldots=0,1,2$ correspond to the time and space directions on the 3D Minkowskian spacetime manifold and the Latin indices $i, j, k \ldots=1,2$ stand for the space directions only. We adopt the convention of (i) the left derivative w.r.t. all the fermionic fields of our theory in the computation of the equations of motion, canonical conjugate momenta, Noether conserved currents, etc., and (ii) the derivative w.r.t. the antisymmetric tensor field $B_{\mu \nu}$ as: $\left(\partial B_{\rho \sigma} / \partial B_{\mu \nu}\right)=\frac{1}{2!}\left(\delta_{\rho}^{\mu} \delta_{\sigma}^{\nu}-\delta_{\sigma}^{\mu} \delta_{\rho}^{\nu}\right)$, etc. The convention for the 3D Levi-Civita tensor is chosen to be such that: $\varepsilon_{012}=+1=\varepsilon^{012}$ and other relationships are: $\varepsilon^{\mu \nu \lambda} \varepsilon_{\mu \nu \lambda}=3!, \quad \varepsilon^{\mu \nu \lambda} \varepsilon_{\mu \nu p}=2!\delta_{\rho}^{\lambda}$, etc. We denote the off-shell nilpotent (anti-)BRST symmetry transformations by the symbols $s_{(a) b}$ and corresponding {\it Noether}
conserved charges carry the symbols $Q_{(a) b}$. Being fermionic in nature, the (anti-)BRST transformation operators $s_{(a) b}$ commute with all the bosonic fields of our theory and they anticommute with the fermionic fields. The notation overdot (i.e. $\dot{\Phi}$) on a generic field $\Phi$ in our theory has been used occasionally to denote the partial derivative w.r.t. time (i.e. $\dot{\Phi}=\partial_{0} \Phi \equiv \partial \Phi / \partial t$) in the natural units where $\hbar=c=1$. In other words, ultimately, we have $\partial_{0}=(\partial / \partial t)$. \\

\vskip0.7cm

%2

\section{Preliminary: Gauge Symmetry Transformations}

Our present section is divided into three concise subsections. In Subsec. 2.1, we derive the Noether conserved current and the corresponding conserved charge from the infinitesimal gauge symmetry transformations. Our Subsec. 2.2 is devoted to the discussion on the
constraint analysis where we show the existence of the first-class constraints on our theory in the terminology of Dirac's prescription for the classification scheme of constraints [37-41]. Finally, in Sebsec. 2.3, we establish a deep connection between the above conserved Noether charge and the first-class constraints (that exist on our theory).\\

\vskip0.7cm

\subsection{Noether Current and Charge: Continuous Symmetry}

We begin with the Lagrangian density $\left(\mathcal{L}_{(0)}^{(M S)}\right)$ for the three $(2+1)$-dimensional massive Abelian 2-form theory, with the rest mass $m$ for the 2-form field, as follows (see, e.g. [26])
\begin{eqnarray}
\mathcal{L}_{(0)}^{(M S)}&=& \frac{1}{2}\left(H_{012}+m \tilde{\varphi}\right)^{2}-\frac{m^{2}}{4} B_{\mu \nu} B^{\mu \nu}-\frac{1}{4} \Sigma^{\mu \nu} \Sigma_{\mu \nu} \nonumber\\
&+&\frac{m}{2} B^{\mu \nu}\left(\Sigma_{\mu \nu}+\varepsilon_{\mu \nu \lambda} \partial^{\lambda} \tilde{\varphi}\right)
-\frac{1}{2} \partial_{\mu} \tilde{\varphi} \partial^{\mu} \tilde{\varphi}, 
\end{eqnarray}
which has been derived from the original Lagrangian density $\left(\mathcal{L}_{(0)}\right)$ for the massive Abelian 2-form $\left[B^{(2)}=\frac{1}{2!} B_{\mu \nu}\left(d x^{\mu} \wedge d x^{\nu}\right)\right]$ field $B_{\mu \nu}$ with (i) the totally antisymmetric field strength tensor $H_{\mu \nu \lambda}$, and (ii) the rest mass $m$ (see, e.g. [26] for details)
\begin{eqnarray}
\mathcal{L}_{(0)}=\frac{1}{12} H^{\mu \nu \lambda} H_{\mu \nu \lambda}-\frac{m^{2}}{4} B^{\mu \nu} B_{\mu \nu}, 
\end{eqnarray}
by using the modified version of the standard St${\ddot u}$ckelberg technique of replacement in 3D
\begin{eqnarray}
B_{\mu \nu} \longrightarrow B_{\mu \nu}-\frac{1}{m} \Sigma_{\mu \nu}-\frac{1}{m} \varepsilon_{\mu \nu \lambda} \partial^{\lambda} \tilde{\varphi}, 
\end{eqnarray}
where $\tilde{\varphi}$ is the pseudo-scalar field and $\varepsilon_{\mu \nu \lambda}$ is the totally antisymmetric 3D Levi-Civita tensor where we have assumed that there is no parity violation in our theory. The superscript ($MS$) on the above Lagrangian density (1) denotes that it has been obtained from (2) by taking into account the modified version of the St${\ddot u}$ckelberg formalism (3). In the equations (1) and (3), we have the field strength tensor $\Sigma_{\mu \nu}=\partial_{\mu} \phi_{\nu}-\partial_{\nu} \phi_{\mu}$ for the St${\ddot u}$ckelberg Lorentz vector field $\phi_{\mu}$ which is derived from the 2-form $\left[\Sigma^{(2)}=d \Phi^{(1)} \equiv \frac{1}{2!} \Sigma_{\mu \nu}\left(d x^{\mu} \wedge d x^{\nu}\right)\right]$ where $d=\partial_{\mu} d x^{\mu}$ [with $\left.d^{2}=\frac{1}{2!}\left(\partial_{\mu} \partial_{\nu}-\partial_{\nu} \partial_{\mu}\right)\left(d x^{\mu} \wedge d x^{\nu}\right)=0\right]$ is the exterior derivative of differential geometry (see, e.g. [21-24] for details) and the 1-form $\Phi^{(1)}=\phi_{\mu} d x^{\mu}$ defines the St${\ddot u}$ckelberg Lorentz vector field $\phi_{\mu}$. It is the peculiarity of the 3D Minkowskian flat spacetime that (i) the field strength tensor $H_{\mu \nu \lambda}=\partial_{\mu} B_{\nu \lambda}+\partial_{\nu} B_{\lambda \mu}+\partial_{\lambda} B_{\mu \nu}$ (derived from the 3 -form $\left.\left[H^{(3)}=d B^{(2)} \equiv \frac{1}{3!} H_{\mu \nu \lambda}\left(d x^{\mu} \wedge d x^{\nu} \wedge d x^{\lambda}\right)\right]\right)$ has only a single independent component which is nothing but $H_{012}$, (ii) the kinetic term for the gauge field becomes: $\frac{1}{12} H^{\mu \nu \lambda} H_{\mu \nu \lambda}=\frac{1}{2} H^{012} H_{012} \equiv \frac{1}{2} H_{012} H_{012}$, and (iii) the component $H_{012}$ changes under the substitution (3) which, ultimately, leads to the derivation 
[26] of the first term of the Lagrangian density (1). One of the key observations is the fact that the pseudo-scalar field appears in the theory with the negative kinetic term. However, it is endowed with a well-defined rest mass $m$ as has been shown in our earlier works (see, e.g. [14,25]). The other crucial observation is the appearance of the higher derivative terms (apart from the usual second order derivative terms) due to the modified version of the replacement (3). However, we have used the simple theoretical trick of the on-shell condition\footnote{We have been able to apply the same trick in the context of 
4D modified massive Abelian 2-form theory [25] to obtain the well-defined 
{\it coupled} Lagrangian densities. We have established that the higher derivative terms are 
{\it not} useless because they lead to the derivations of the correct and appropriate terms of the Lagrangian densities which were incorporated into the
{\it coupled} Lagrangian densities of our earlier works  (on the above 2D and 4D theories [15,14,25]) by trial and error method.}
to replace the higher derivative terms in terms of mass term to obtain the well-defined 3D Lagrangian density (1) for the modified 3D massive Abelian 2-form theory without any higher derivative terms.

The Lagrangian density (1) respects the infinitesimal, local and continuous gauge symmetry transformations. These transformations are as follows, namely;
\begin{eqnarray}\label{4}
\delta_{g} B_{\mu \nu}=-\left(\partial_{\mu} \Lambda_{\nu}-\partial_{\nu} \Lambda_{\mu}\right), & \delta_{g} \phi_{\mu}=\left(\partial_{\mu} \Lambda-m \Lambda_{\mu}\right), \quad \delta_{g} H_{\mu \nu \lambda}=0 \nonumber\\
\delta_{g} \Sigma_{\mu \nu}=-m\left(\partial_{\mu} \Lambda_{\nu}-\partial_{\nu} \Lambda_{\mu}\right), & \delta_{g} \tilde{\varphi}=0, \quad \delta_{g} \mathcal{L}_{(0)}^{(MS)}=\partial_{\mu}\left[-m \varepsilon^{\mu \nu \sigma} \Lambda_{\nu} \partial_{\sigma} \tilde{\varphi}\right],
\end{eqnarray}
which demonstrate that the action integral, corresponding to the Lagrangian density $\mathcal{L}_{(0)}^{(M S)}$, remains invariant under the above gauge symmetry transformations for the physical fields that vanish off as $x \rightarrow \pm \infty$. In equation (4), the Lorentz vector $\Lambda_\mu$ and
Lorentz scalar $\Lambda$ are the local and infinitesimal gauge symmetry transformation parameters. The {\it latter} parameter is present in (4)
due to the stage-one reducibility in our theory.
According to celebrated Noether's theorem, the above gauge symmetry transformations lead to the derivation of the Noether current $J^{\mu}$ as
\begin{eqnarray}
J^{\mu} & =\left(\delta_{g} \tilde{\varphi}\right) \frac{\partial \mathcal{L}_{(0)}^{(M S)}}{\partial\left(\partial_{\mu} \tilde{\varphi}\right)}+\left(\delta_{g} \phi_{\rho}\right) \frac{\partial \mathcal{L}_{(0)}^{(M S)}}{\partial\left(\partial_{\mu} \phi_{p}\right)}+\left(\delta_{g} B_{\rho \sigma}\right) \frac{\partial \mathcal{L}_{(0)}^{(M S)}}{\partial\left(\partial_{\mu} B_{\rho \sigma)}\right.}+m \varepsilon^{\mu \nu \sigma} \Lambda_{\nu} \partial_{\sigma} \tilde{\varphi} \nonumber\\
& \equiv\left(m B^{\mu \nu}-\Sigma^{\mu \nu}\right)\left(\partial_{\nu} \Lambda-m \Lambda_{\nu}\right)-\varepsilon^{\mu \nu \sigma}\left(\partial_{\nu} \Lambda_{\sigma}\right)\left(H_{012}+m \tilde{\varphi}\right)+m \varepsilon^{\mu \nu \sigma} \Lambda_{\nu} \partial_{\sigma} \tilde{\varphi}, 
\end{eqnarray}
where we have used: $H_{012}=\frac{1}{2} \varepsilon^{\mu \nu \sigma} \partial_{\mu} B_{\nu \sigma}$. 
The conservation law (i.e. $\partial_{\mu} J^{\mu}=0$) can be proven, in a straightforward manner, by 
using the following EL-EoMs that can be readily derived from the Lagrangian density (1) w.r.t. the 
basic fields $\tilde{\varphi}, \phi_{\mu}$ and $B_{\mu \nu}$, namely;
\begin{eqnarray}
\left(\square+m^{2}\right) \tilde{\varphi}=0, \quad \partial_{\mu}\left(m B^{\mu \nu}-\Sigma^{\mu \nu}\right)=0, \quad 
\varepsilon^{\mu \nu \sigma} \partial_{\sigma} H_{012}=m\left(\Sigma^{\mu \nu}-m B^{\mu \nu}\right).
\end{eqnarray}
According to the basic tenets behind the Noether theorem, we obtain the explicit expression for the conserved charge $Q$, from the above Noether conserved current (5), as follows
\begin{eqnarray}
Q=\int d^{2} x J^{0} & \equiv \int d^{2} x\left[\left(m B^{0 \nu}-\Sigma^{0 \nu}\right)\left(\partial_{\nu} \Lambda-m \Lambda_{\nu}\right)\right. \nonumber\\
& \left.\left.-\varepsilon^{0 \nu \sigma}\left(\partial_{\nu} \Lambda_{\sigma}\right)\left(H_{012}+m \tilde{\varphi}\right)+m \varepsilon^{0 \nu \sigma} 
\Lambda_{\nu} \partial_{\sigma} \tilde{\varphi}\right)\right],
\end{eqnarray}
which turns out to be the generator for the gauge symmetry transformations in (4) if we use the appropriate non-vanishing canonical commutators for our theory. We do this exercise in our Subsect. 2.3 (see below) where a connection between the first-class constraints of our theory and the above conserved charge is established.

We end this subsection with a couple of remarks. First of all, we note that the pseudo-scaler (PS) field $\tilde \varphi $ 
is endowed with the negative kinetic term which emerges due to the modified version of the St{\" u}ckelberg formalism [cf. Eq. (3)]. 
Second, this PS field obeys the Klein-Gordon equation of motion [cf. Eq. (6)] which is derived from the Lagrangian density (1).
 These two observations establish that (i) the PS field 
is an ``exotic'' field because it carries the {\it negative} kinetic term, and (ii) the PS field is relativistic and it is
endowed with a rest mass $m$ equal to the rest mass of the Abelian 2-form gauge field [cf. Eq. (2)].  \\

\vskip 1.8cm

\subsection{Constraint Analysis: First-Class Constraints}

We have a set of continuous, local and infinitesimal
gauge symmetry transformations (4) which is respected by the action integral corresponding to the Lagrangian density (1) of our St{\" u}ckelberg-modified 
3D massive Abelian 2-form theory. The existence of the local gauge symmetry transformations (in the context of a physical system) 
{\it always} owes its origin to the presence of a set of first-class constraints (on that physical system). To corroborate this statement, we perform the constraint analysis of our present modified 3D massive Abelian 2-form theory. Towards this goal in mind, first of all, we note that the explicit expressions for the canonical conjugate momenta [derived from the Lagrangian density (1)] w.r.t. the basic fields $\tilde{\varphi}, \phi_{\mu}$ and $B_{\mu \nu}$ are
\begin{eqnarray}
&&\Pi_{(\tilde{\varphi})}  =\frac{\partial \mathcal{L}_{(0)}^{(M S)}}{\partial\left(\partial_{0} \tilde{\varphi}\right)}=-\dot{\varphi}+\frac{m}{2} \varepsilon^{0 i j} B_{i j}, \quad \Pi_{(\phi)}^{\mu}=\frac{\partial \mathcal{L}_{(0)}^{(M S)}}{\partial\left(\partial_{0} \phi_{\mu}\right)}=m B^{0 \mu}-\Sigma^{0 \mu}, \nonumber\\
&&\Pi_{(B)}^{\mu \nu}  =\frac{\partial \mathcal{L}_{(0)}^{(M S)}}{\partial\left(\partial_{0} B_{\mu \nu}\right)}=\frac{1}{2} \varepsilon^{0 \mu \nu}\left(H_{012}+m \tilde{\varphi}\right),
\end{eqnarray}
which demonstrate that we have the following primary constraints on our theory, namely;
\begin{eqnarray}
\Pi_{(\phi)}^{0}=m B^{00}-\Sigma^{00} \approx 0, \quad \Pi_{(B)}^{0 i}=\frac{1}{2} \varepsilon^{00 i}\left(H_{012}+m \tilde{\varphi}\right) \approx 0,
\end{eqnarray}
where we have used Dirac's notation for the weakly zero (i.e. $\approx 0$) which implies that we are allowed to take a first-order time derivative on the above primary constraints. The above constraints are nothing but the components (i.e. $\Pi_{(\phi)}^{0}, \Pi_{(B)}^{0 i}$) of the conjugate momenta w.r.t. the 
St${\ddot u}$ckelberg  vector field $\phi_{\mu}$ and the massive antisymmetric gauge field $B_{\mu \nu}$, respectively. It is worthwhile to point out that $\Pi_{(B)}^{00}=0$ is {\it strongly} equal to zero and it is not a constraint on our present theory. The subscripts $[(\tilde{\varphi}),(\phi),(B)]$ on the canonical conjugate momenta in (8) denote that these are defined w.r.t. the basic fields $\tilde{\varphi}, \phi_{\mu}$ and $B_{\mu \nu}$ of our theory from the starting Lagrangian density (1), respectively. The requirement of the time-evolution invariance of the above primary constraints (PCs) leads to the secondary constraints on our theory. The most appropriate approach, to obtain the time-evolution invariance of the PCs, is the Hamiltonian formalism. However, for the simple system like our present modified 3D massive Abelian 2-form theory, the EL-EoMs of the theory, derived from the starting Lagrangian density (1), are good enough (see, e.g. [42] for details).

To corroborate the above statement, we focus on the last two entries of the EL-EoMs (6). First of all, we note that, from the 
{\it second} entry of (6), we have the following equation of motion for the choice $\nu=0$, namely;
\begin{eqnarray} 
\partial_{0}\left(m B^{00}-\Sigma^{00}\right)+\partial_{i}\left(m B^{i 0}-\Sigma^{i 0}\right)=0 \quad \Longrightarrow \quad \partial_{0} \Pi_{(\phi)}^{0}=\partial_{i} \Pi_{(\phi)}^{i} \approx 0,
\end{eqnarray}
which captures the time-evolution invariance of the primary constraint $\Pi_{\{\phi\rangle}^{0} \approx 0$ where $\Pi_{(\phi)}^{i}=m B^{0 i}-\Sigma^{0 i}$ is the space component of the momenta w.r.t. the St${\ddot u}$ckelberg  vector field $\phi_{\mu}[$ cf. Eq. (8)]. Thus, we have obtained one of the secondary constraints of our theory as: $\partial_{i} \Pi_{(\phi)}^{i} \approx 0$ (w.r.t. the primary constraint $\Pi_{(\phi)}^{0} \approx 0$ [cf. Eq. (9)]). It can be readily checked that the primary constraints in (9) commute with {\it this} secondary constraint because 
{\it all} are the components of the conjugate momenta [cf. Eq. (8)] and a space derivative on one of them. Now we are in the position to concentrate on the last entry of (6) which has been actually derived from the following EL-EoM w.r.t. the massive gauge field $B_{\mu \nu}$ [from the Lagrangian density (1)], namely;
\begin{eqnarray}
\partial_{\lambda}\left(\frac{1}{2} \varepsilon^{\mu \nu \lambda}\left[H_{012}+m \tilde{\varphi}\right]\right)=\frac{m}{2}\left(\Sigma^{\mu \nu}-m B^{\mu \nu}\right)+\frac{m}{2} \varepsilon^{\mu \nu \lambda} \partial_{\lambda} \tilde{\varphi}.
\end{eqnarray}
If we make the choice: $\mu=0, \nu=i$, we obtain the following from the above equation
\begin{eqnarray}
\partial_{0} \Pi_{(B)}^{0 i}=\frac{m}{2} \Pi_{(\phi)}^{i}-\partial_{j} \Pi_{(B)}^{j i}
-\frac{m}{2} \varepsilon^{0 i j} \partial_{j} \tilde{\varphi} \approx 0,
\end{eqnarray}
which is nothing but the requirement of the time-evolution invariance of the primary constraint $\Pi_{(B)}^{0 i} \approx 0$
[cf. Eq. (9)]. In other words, we have derived the secondary constraint (w.r.t. the primary constraint $\Pi_{(B)}^{0 i} \approx 0$ [cf. Eq. (9)]) on our theory as
\begin{eqnarray}
\frac{m}{2} \Pi_{(\phi)}^{i}-\partial_{j} \Pi_{(B)}^{j i}-\frac{m}{2} \varepsilon^{0 i j} \partial_{j} \tilde{\varphi} \approx 0,
\end{eqnarray}
where the space components of the canonical conjugate momenta are: $\Pi_{(\phi)}^{i}=m B^{0 i}-\Sigma^{0 i}$ and $\Pi_{(B)}^{i j}=\frac{1}{2} \varepsilon^{0 i j}\left(H_{012}+m \tilde{\varphi}\right)$ [cf. Eq. (8)]. It is straightforward to note that all the primary and secondary constraints of our theory commute among themselves. Hence, they belong to the first-class category of constraints (according to Dirac's prescription for the classification scheme of constraints [37-41]). Ultimately, we have two sets of primary and secondary constraints on our modified 3D Massive Abelian 2 -form theory which are:
\begin{eqnarray}
&&\Pi_{(\phi)}^{0}  =m B^{00}-\Sigma^{00} \approx 0, \quad \Pi_{(B)}^{0 i}=\frac{1}{2} \varepsilon^{00 i}\left(H_{012}+m \tilde{\varphi}\right) \approx 0, \nonumber\\
&&\partial_{i} \Pi_{(\phi)}^{i}  \approx 0, \quad \frac{m}{2} \Pi_{(\phi)}^{i}-\partial_{j} \Pi_{(B)}^{j i}-\frac{m}{2} \varepsilon^{0 i j} \partial_{j} \tilde{\varphi} \approx 0.
\end{eqnarray}
It is straightforward to check that all the above constraints commute among themselves. Hence, we have a set of four first-class constraints on our theory.

We conclude this subsection with the a couple of final remarks. First, there is no primary constraint associated with the canonical conjugate momentum 
$\Pi_{(\tilde {\varphi})}$ because it is not strongly/weakly equal to zero. Second, to be precise, the the exact number of constraints in (14) are eight (if we take into account the components). However, we shall stick with calling {\it four} constraints only for the sake of brevity.\\

\subsection{Noether Conserved Charge, Gauge Symmetry Transformations and First-Class Constraints: A Deep Relationship}

The central purpose of our present subsection is to establish that the conserved Noether charge (7) is the generator for the infinitesimal gauge symmetry transformations (4) of our theory. Since there are 
{\it four} first-class constraints on our theory [cf. Eq. (14)], we have to be careful in expanding the r.h.s of the conserved Noether charge (7) so that the first-class constraints are {\it not} strongly set equal to zero. In other words, we have the following:
\begin{eqnarray}
& Q=\int d^{2} x\left[\left(m B^{00}-\Sigma^{00}\right)\left(\partial_{0} \Lambda-m \Lambda_{0}\right)+\left(m B^{0 i}-\Sigma^{0 i}\right)\left(\partial_{i} \Lambda-m \Lambda_{i}\right)+m \varepsilon^{0 i j} \Lambda_{i} \partial_{j} \tilde{\varphi}\right. \nonumber\\
& \left.-\frac{1}{2} \varepsilon^{00 i}\left(\partial_{0} \Lambda_{i}-\partial_{i} \Lambda_{0}\right)\left(H_{012}+m \tilde{\varphi}\right)-\frac{1}{2} \varepsilon^{0 i j}\left(\partial_{i} \Lambda_{j}-\partial_{j} \Lambda_{i}\right)\left(H_{012}+m \tilde{\varphi}\right)\right].
\end{eqnarray}
Using the definitions of the canonical conjugate momenta from Eq. (8), we can express the above charge in terms of the precise components of the momenta as follows:
\begin{eqnarray}
Q & =\int d^{2} x\left[\Pi_{(\phi)}^{0}\left(\partial_{0} \Lambda-m \Lambda_{0}\right)+\Pi_{(\phi)}^{i}\left(\partial_{i} \Lambda-m \Lambda_{i}\right)\right]+m \varepsilon^{0 i j} \Lambda_{i} \partial_{j} \tilde{\varphi}\nonumber\\
& -\Pi_{(B)}^{0 i}\left(\partial_{0} \Lambda_{i}-\partial_{i} \Lambda_{0}\right)-\Pi_{(B)}^{i j}\left(\partial_{i} \Lambda_{j}-\partial_{j} \Lambda_{i}\right).
\end{eqnarray}
At this juncture, we are in the position to define the non-vanishing canonical commutators for our 3D 
{\it modified} massive theory (in the natural units where $\hbar=c=1$ ) as 
\begin{eqnarray}
%$$
%\begin{aligned}
&&{\left[\phi_{0}(\vec{x}, t), \Pi_{(\phi)}^{0}(\vec{y}, t)\right] }  =i \delta^{(2)}(\vec{x}-\vec{y}),\nonumber\\
&&{\left[\phi_{i}(\vec{x}, t), \Pi_{(\phi)}^{j}(\vec{y}, t)\right] }  =i \delta_{i}^{j} \delta^{(2)}(\vec{x}-\vec{y}),\nonumber\\
%\end{aligned}
%$$
&&{\left[B_{0 i}(\vec{x}, t), \Pi_{(B)}^{0 j}(\vec{y}, t)\right] }  =i \delta_{i}^{j} \delta^{(2)}(\vec{x}-\vec{y}), \nonumber\\
&&{\left[B_{i j}(\vec{x}, t), \Pi_{(B)}^{k l}(\vec{y}, t)\right] }  =\frac{i}{2!}\left(\delta_{i}^{k} \delta_{j}^{l}-\delta_{i}^{l} \delta_{j}^{k}\right) \delta^{(2)}(\vec{x}-\vec{y}),\nonumber\\
&&{\left[\tilde{\varphi}(\vec{x}, t), \Pi_{(\tilde{\varphi})}(\vec{y}, t)\right] }  =i \delta^{(2)}(\vec{x}-\vec{y}),
\end{eqnarray}
where the above brackets are known as the equal-time canonical commutators. All the rest of the equal-time canonical commutators, as per the rules of the canonical quantization scheme, are equal to zero. Using the non-vanishing canonical commutators (17), we observe that the Noether conserved charge (16) is the generator for the infinitesimal gauge transformations (4) as is evident from the following
\begin{eqnarray}
\delta_{g} \phi_{0}(\vec{x}, t) & =&-i\left[\phi_{0}(\vec{x}, t), Q\right]=\left(\partial_{0} \Lambda-m \Lambda_{0}\right),\nonumber\\
\delta_{g} \phi_{i}(\vec{x}, t) & =&-i\left[\phi_{i}(\vec{x}, t), Q\right]=\left(\partial_{i} \Lambda-m \Lambda_{i}\right), \nonumber\\
\delta_{g} B_{0 i}(\vec{x}, t) & =&-i\left[B_{0 i}(\vec{x}, t), Q\right]=-\left(\partial_{0} \Lambda_{i}-\partial_{i} \Lambda_{0}\right), \nonumber\\
\delta_{g} B_{i j}(\vec{x}, t) & =&-i\left[B_{i j}(\vec{x}, t), Q\right]=-\left(\partial_{i} \Lambda_{j}-\partial_{j} \Lambda_{i}\right), \nonumber\\
\delta_{g} \tilde{\varphi}(\vec{x}, t) & =&-i[\tilde{\varphi}(\vec{x}, t), Q]=0.
\end{eqnarray}
where the last entry is correct because we do not have the canonical conjugate momentum $\Pi_{(\tilde{\varphi})}$, corresponding to the pseudo-scalar field $\tilde{\varphi}$, in the expression for the Noether conserved charge $Q$. In the above equation (18), the covariant form of the infinitesimal gauge 
symmetry transformations (4), have been expressed in their component forms.

We would like to say, at this stage, a few words about the connection between the Noether conserved charge $Q$ [cf. Eq. (16)] and the first-class constraints (14) that exist on our theory. It is straightforward to note that we can throw away the total space derivative terms from the expression for $Q$ in (16) by exploiting the mathematical potential and physical arguments of Gauss's divergence theorem. In other words, we take into account the following inputs due to the above celebrated theorem, namely;
\begin{eqnarray}
&& \int d^{2} x \Pi_{(\phi)}^{i}\left(\partial_{i} \Lambda\right)=-\int d^{2} x\left(\partial_{i} \Pi_{(\phi)}^{i}\right) \Lambda, \nonumber\\
&& -\int d^{2} x \Pi_{(B)}^{i j}\left(\partial_{i} \Lambda_{j}-\partial_{j} \Lambda_{i}\right)=+\int d^{2} x\left[\left(\partial_{i} \Pi_{(B)}^{i j}\right) \Lambda_{j}+\left(\partial_{j} \Pi_{(B)}^{j i}\right) \Lambda_{i}\right],
\end{eqnarray}
in addition to the simple mathematical tricks of re-writing the expressions
\begin{eqnarray}
 -m \Pi_{(\phi)}^{i} \Lambda_{i} &=&-\frac{m}{2} \Pi_{(\phi)}^{i} \Lambda_{i}-\frac{m}{2} \Pi_{(\phi)}^{j} \Lambda_{j}, \nonumber\\
 +m \varepsilon^{0 i j} \Lambda_{i} \partial_{j} \tilde{\varphi} &=& +\frac{m}{2} \varepsilon^{0 i j} \Lambda_{i} \partial_{j} \tilde{\varphi}+\frac{m}{2} \varepsilon^{0 j i} \Lambda_{j} \partial_{i} \tilde{\varphi}, 
\end{eqnarray}
to re-express the charge $Q$ in (16) as follows:
\begin{eqnarray}
&&Q  =\int d^{2} x\left[\Pi_{(\phi)}^{0}\left(\partial_{0} \Lambda-m \Lambda_{0}\right)-\left(\partial_{i} \Pi_{(\phi)}^{i}\right) \Lambda-\Pi_{(B)}^{0 i}\left(\partial_{0} \Lambda_{i}-\partial_{i} \Lambda_{0}\right)\right) \nonumber\\
&& -\left(\frac{m}{2} \Pi_{(\phi)}^{i}-\partial_{j} \Pi_{(B)}^{j i}-\frac{m}{2} \varepsilon^{0 i j} \partial_{j} \tilde{\varphi}\right) \Lambda_{i}
 \left.
 -\left(\frac{m}{2} \Pi_{(\phi)}^{j}-\partial_{i} \Pi_{(B)}^{i j}-\frac{m}{2} \varepsilon^{0 j i} \partial_{i} \tilde{\varphi}\right) \Lambda_{j}\right]. 
\end{eqnarray}
In the above, we have also used the antisymmetric (i.e. $\Pi_{(B)}^{\mu \nu}=-\Pi_{(B)}^{\nu \mu}$) property of the conjugate momenta (i.e. $\Pi_{(B)}^{0 i}=-\Pi_{(B)}^{i 0}$, $\Pi_{(B)}^{i j}=-\Pi_{(B)}^{j i}$) w.r.t. the antisymmetric tensor gauge field $B_{\mu \nu}$. A close look at the expression for the gauge symmetry generator $Q$ in (21) demonstrates that we have been able to express {\it it}
 in terms of the primary and secondary constraints that have been listed in (14). In more sophisticated language, the infinitesimal gauge symmetry transformations (4) are generated by the first-class constraints [cf. Eq. (14)] in the terminology of Dirac's prescription for the classification scheme of constraints [37-41]. These constraints are present in the generator $Q$ [cf. Eq. (21)]. Finally, we would like to add that the above expression for the conserved Noether charge matches with the standard formula for the generator of the classical gauge symmetry transformations that 
has been obtained (see, e.g. [43] for details) in terms of the first-class constraints.\\

%3
\section{BRST Transformations: Lagrangian Density and Nilpotency Property of the BRST Charge}

The theoretical contents of this section are divided into two parts. In Subsec. 3.1, we derive the Noether conserved charge $Q_{b}$ (from the conserved Noether current) and show that (i) it is the generator for the nilpotent BRST symmetry transformations (26), and (ii) it is 
{\it not} nilpotent (i.e. $Q_{b}^{2} \neq 0$) of order two because of the presence of the non-trivial CF-type restrictions on our theory (see, e.g. [45] for details). Our Subsec. 3.2 is devoted to the derivation of the nilpotent $\left(Q_{B}^{2}=0\right)$ version of the BRST charge $Q_{B}$ from the  non-nilpotent version of the Noether charge $Q_{b}$.\\

\subsection{Noether Conserved Charge: BRST Symmetries}

In this subsection, first of all, we focus on the derivation of the BRST-invariant Lagrangian density which is a generalization of the classical 
St${\ddot u}$ckelberg-modified classical Lagrangian density (1) to its counterpart quantum version that incorporates (i) the gauge-fixing terms, and (ii) the Faddeev-Popov (FP) ghost terms. In this connection, we would like to mention that the properly gauge-fixed Lagrangian density\footnote{This gauge-fixed Lagrangian density (22) is such that the scalar and pseudo-scalar fields obey the Klein-Gordon equations of motion [i.e. $\left.\left(\square+m^{2}\right) \phi=0,\left(\square+m^{2}\right) \tilde{\varphi}=0\right]$ despite the fact that they are endowed with the kinetic terms that carry 
explicitly the opposite signs.}
 for our present system of the modified massive 3D Abelian 2-form gauge theory has been written in our earlier work [26]
\begin{eqnarray}
{\cal L}_{(0)}^{(MS)}+{\cal L}_{(g f)}^{(B)} & =&\frac{1}{2}\left(H_{012}+m \tilde{\varphi}\right)^{2}-\frac{m^{2}}{4} B^{\mu \nu} B_{\mu \nu}+\frac{m}{2} B^{\mu \nu}\left[\Sigma_{\mu \nu}+\varepsilon_{\mu \nu \sigma} \partial^{\sigma} \tilde{\varphi}\right]-\frac{1}{4} \Sigma^{\mu \nu} \Sigma_{\mu \nu} \nonumber\\
& -&\frac{1}{2} \partial_{\mu} \tilde{\varphi} \partial^{\mu} \tilde{\varphi}-\frac{1}{2}(\partial \cdot \phi+m \phi)^{2}+\frac{1}{2}\left(\partial^{\nu} B_{\nu \mu}-\partial_{\mu} \phi+m \phi_{\mu}\right)^{2},
\end{eqnarray}
where ${\cal L}_{(g f)}^{(B)}$ is the gauge-fixing part of the Lagrangian density of our theory. In this term, the appropriate mass dimensions (in the natural units) have been taken into account as far as the proper gauge-fixing terms for the St${\ddot u}$ckelberg vector field $\phi_{\mu}$ and the massive gauge field $B_{\mu \nu}$ are concerned. The gauge-fixing term for the vector field $\phi_{\mu}$ corresponds to the famous 't Hooft gauge that has been invoked in the context of the quantizations
of the Abelian Higgs model and modified massive Abelian 1-form theories (see, e.g. [44] for details). The superscript $(B)$ on the Lagrangian density 
${\cal L}_{(g f)}^{(B)}$ denotes that we are dealing with the modified massive Abelian 2-form gauge theory where the basic field is the antisymmetric tensor massive gauge field $B_{\mu \nu}$.

All the quadratic terms, in the above gauge-fixed Lagrangian density (22), can be linearized by invoking the Nakanishi-Lautrup type 
auxiliary fields\footnote{It will be noted that our present choices of the signs for the Nakanishi-Lautrup type auxiliary fields, in the linearization processes of the gauge-fixing terms, are different from our earlier work. These differences can be seen in our equations (23) and (48) and the corresponding equations in our earlier work [45].}
as follows:
\begin{eqnarray}
&& \mathcal{L}_{(b)}=\mathcal{B}\left(\frac{1}{2} \varepsilon^{\mu \nu \sigma} \partial_{\mu} B_{\nu \sigma}+m \tilde{\varphi}\right)-\frac{\mathcal{B}^{2}}{2}-\frac{m^{2}}{4} B^{\mu \nu} B_{\mu \nu}+\frac{m}{2} B^{\mu \nu}\left(\Sigma_{\mu \nu}+\varepsilon_{\mu \nu \sigma} \partial^{\sigma} \tilde{\varphi}\right)+\frac{B^{2}}{2}  \nonumber\\
&& -B(\partial \cdot \phi+m \phi)-\frac{1}{4} \Sigma^{\mu \nu} \Sigma_{\mu \nu}-\frac{1}{2} \partial_{\mu} \tilde{\varphi} \partial^{\mu} \tilde{\varphi}+B^{\mu}\left(\partial^{\nu} B_{\nu \mu}-\partial_{\mu} \phi+m \phi_{\mu}\right)-\frac{B^{\mu} B_{\mu}}{2}.
\end{eqnarray}
In the above, we have utilized the auxiliary field $\mathcal{B}$ to linearize the kinetic term where we have taken into account the covariant form of $H_{012}$ as: $H_{012}=\frac{1}{2} \varepsilon^{\mu \nu /} \partial_{\mu} B_{\nu \sigma}$. Similarly, the auxiliary fields $B$ and $B_{\mu}$ have been invoked to linearize the quadratic gauge-fixing terms for the St${\ddot u}$ckelberg vector field $\phi_{\mu}$ and gauge field $B_{\mu \nu}$, respectively. It is worth pointing out that we have a scalar field, too, in our theory which is due to the reducibility property of the massive gauge field $B_{\mu \nu}$. The FP-ghost part of the BRST-invariant Lagrangian density for our massive Abelian 2-form theory is as follows\footnote{The systematic derivation of the gauge-fixing and FP-ghost terms for the modified massive Abelian 2-form theory has been performed in our earlier work (see, e.g. [14] for details) where this theory has been proven to be a model for Hodge theory. In our present endeavor, we have taken into account the simplest form of the FP-ghost Lagrangian density which has been considered in our another earlier work (see, e.g. [45]) where the proof of this theory to be an example of Hodge theory has {\it not} been given any importance.}
(see, e.g. [45] for details)
\begin{eqnarray}
\mathcal{L}_{(F P)} & =&\partial_{\mu} \bar{\beta} \partial^{\mu} \beta-m^{2} \bar{\beta} \beta+\left(\partial_{\mu} \bar{C}_{\nu}-\partial_{\nu} \bar{C}_{\mu}\right)\left(\partial^{\mu} C^{\nu}\right)-\left(\partial_{\mu} \bar{C}-m \bar{C}_{\mu}\right)\left(\partial^{\mu} C-m C^{\mu}\right)  \nonumber\\
& +&(\partial \cdot \bar{C}+\rho+m \bar{C}) \lambda+(\partial \cdot C-\lambda+m C) \rho, 
\end{eqnarray}
where $\left(\bar{C}_{\mu}\right) C_{\mu}$ are the fermionic (i.e. $C_{\mu}^{2}=\bar{C}_{\mu}^{2}=0, C_{\mu} \bar{C}_{\nu}+\bar{C}_{\nu} C_{\mu}=0, C_{\mu} C_{\nu}+C_{\nu} C_{\mu}=$ $0, \bar{C}_{\mu} \bar{C}_{\nu}+\bar{C}_{\nu} \bar{C}_{\mu}=0$) (anti-)ghost fields with ghost numbers $(-1)+1$ and $(\bar{\beta}) \beta$ are the bosonic (anti-)ghost fields with ghost numbers $(-2)+2$, respectively. The pair $(\bar{C}) C$ are the additional set of (anti-)ghost fermionic (i.e. $C^{2}=\bar{C}^{2}=0, C \bar{C}+\bar{C} C=0$) fields with ghost numbers $(-1)+1$. On the other hand, we have $(\rho) \lambda$ as the auxiliary (anti-)ghost fields with ghost numbers $(-1)+1$, respectively, because we note that: $\rho=-\frac{1}{2}(\partial \cdot \bar{C}+m \bar{C})$ and $\lambda= \frac{1}{2}(\partial \cdot C+m C)$. These (anti-)ghost fields are required in the theory to maintain the sacrosanct property of unitarity at any arbitrary order of perturbative computations for a given physical process that is allowed by our BRST quantized theory (see, e.g. [46-48]).

The total Lagrangian density [i.e. ${\cal L}_{B}= {\cal L}_{(0)}^{(M S)}+ {\cal L}_{(g f)}^{(B)}+ {\cal L}_{(F P)}$], which is the sum of equations (23) and (24), transforms to the total spacetime derivative
\begin{eqnarray}
s_{b} \mathcal{L}_{B}=-\partial_{\mu}\left\{\left(\partial^{\mu} C^{\nu}-\partial^{\nu} C^{\mu}\right) B_{\nu}+m \varepsilon^{\mu \nu \sigma} C_{\nu} \partial_{\sigma} \tilde{\varphi}+\lambda B^{\mu}+\rho \partial^{\mu} \beta+\left(\partial^{\mu} C-m C^{\mu}\right) B\right\}, 
\end{eqnarray}
under the following infinitesimal, continuous and off-shell nilpotent (i.e. $s_{b}^{2}=0$) BRST 
symmetry transformations ($(s_{b}$), namely;
\begin{eqnarray}
&& s_{b} B_{\mu \nu}=-\left(\partial_{\mu} C_{\nu}-\partial_{\nu} C_{\mu}\right), \quad s_{b} C_{\mu}=-\partial_{\mu} \beta, \quad s_{b} \bar{C}_{\mu}=-B_{\mu}, \nonumber\\
&& s_{b} \phi_{\mu}=\left(\partial_{\mu} C-m C_{\mu}\right), \quad s_{b} \bar{C}=B, \quad s_{b} C=-m \beta,  \nonumber\\
&& s_{b} \bar{\beta}=-\rho, \quad s_{b} \phi=+\lambda, \quad s_{b} \Sigma_{\mu \nu}=-m\left(\partial_{\mu} C_{\nu}-\partial_{\nu} C_{\mu}\right),  \nonumber\\
&& s_{b}\left[\tilde{\varphi}, \rho, \lambda, \beta, B_{\mu}, B, \mathcal{B}, H_{\mu \nu \lambda}\right]=0.
\end{eqnarray}
As a consequence, the action integral $S=\int d^{3} x \mathcal{L}_{B}$, corresponding to the Lagrangian density $\mathcal{L}_{B}$, remains BRST invariant (i.e. $s_{b} S=0$) for the physical fields that vanish off as $x \rightarrow \pm \infty$ due to Gauss's divergence theorem. Three crucial observations, at this stage, are worth pointing out. First of all, we note that the field-strength tensor $H_{\mu \nu \lambda}$ (owing its origin to the exterior derivative) remains invariant under the BRST symmetry transformations. Second, the 
``exotic'' pseudo-scalar field does not participate in the classical gauge as well as in the BRST symmetry transformations. In other words, it remains inert (i.e. $\delta_{g} \tilde{\varphi}=0, s_{b} \tilde{\varphi}=$ 0) to the classical gauge as well as quantum BRST symmetry transformations [cf. Eqs. (4),(26)]. However, this pseudo-scalar field participates in the (anti-)co-BRST symmetry transformations that have been shown in our earlier works [14-16]. Finally, the above nilpotent BRST symmetry transformations (26) are the generalizations of the gauge
symmetry transformations (4). 
The observation in (25) implies that we can compute the BRST Noether conserved current following the similar kind of formula that is given in (5). We have to take into account the total BRST invariant Lagrangian density which is the sum of (23) and (24) in the application of the analogue of (5). In fact, we obtain the following expression for the BRST current [$J_{(b)}^{\mu}$], namely;
\begin{eqnarray}
J_{(b)}^{\mu} & =& m \varepsilon^{\mu \nu \sigma} C_{\nu} \partial_{\sigma} \tilde{\varphi}+\left(\partial^{\mu} \bar{C}^{\nu}-\partial^{\nu} \bar{C}^{\mu}\right) \partial_{\nu} \beta-\lambda B^{\mu}-\left(\partial^{\mu} C-m C^{\mu}\right) B  \nonumber\\
& -&m\left(\partial^{\mu} \bar{C}-m \bar{C}^{\mu}\right) \beta+\left(m B^{\mu \nu}-\Sigma^{\mu \nu}\right)\left(\partial_{\nu} C-m C_{\nu}\right)-\rho \partial^{\mu} \beta  \nonumber\\
& -&\frac{1}{2}\left[\varepsilon^{\mu \nu \sigma} \mathcal{B}+\left(\eta^{\mu \nu} B^{\sigma}-\eta^{\mu \sigma} B^{\nu}\right)\right]\left(\partial_{\nu} C_{\sigma}-\partial_{\sigma} C_{\nu}\right).
\end{eqnarray}
The conservation law (i.e. $\partial_{\mu} J_{\langle b)}^{\mu}=0$) can be readily proven by using the following EL-EoMs that are derived from the total BRST invariant Lagrangian density, namely;
\begin{eqnarray}
&& \left(\square+m^{2}\right) \beta=0, \quad \partial_{\mu}\left(\partial^{\mu} \bar{C}^{\nu}-\partial^{\nu} \bar{C}^{\mu}\right)=\partial^{\nu} \rho+m\left(\partial^{\nu} \bar{C}-m \bar{C}^{\nu}\right),  \nonumber\\
&& \partial_{\mu}\left(\partial^{\mu} \bar{C}-m \bar{C}^{\mu}\right)=+m \rho, \quad \partial_{\mu}\left(\partial^{\mu} C-m C^{\mu}\right)=-m \lambda,  \nonumber\\
&& \square C_{\mu}-\partial_{\mu}(\partial \cdot C)+\partial_{\mu} \lambda=+m\left(\partial_{\mu} C-m C_{\mu}\right), \quad\left(\square+m^{2}\right) \bar{\beta}=0.
\end{eqnarray}
In addition to the above EL-EoMs (that emerge out from the ghost-sector of the total BRST invariant Lagrangian density), we have to use the following EL-EoMs that emerge out from the non-ghost sector of our theory, namely;
\begin{eqnarray}
&& \partial_{\mu}\left(m B^{\mu \nu}-\Sigma^{\mu \nu}\right)=m B^{\nu}+\partial^{\nu} B, \quad \partial_{\mu} B^{\mu}=m B, \nonumber\\
&& \varepsilon^{\mu \nu \sigma} \partial_{\mu} \mathcal{B}+\left(\partial^{\nu} B^{\sigma}-\partial^{\sigma} B^{\nu}\right)=-m\left(m B^{\nu \sigma}-\Sigma^{\nu \sigma}\right)+m \varepsilon^{\nu \sigma \eta} \partial_{\eta} \tilde{\varphi}.
\end{eqnarray}
Following the basic tenets of the Noether theorem, we can derive the BRST charge $Q_{b}=$ $\int d^{2} x J_{(b)}^{0}$ which will also be conserved because it is derived from the conserved Noether BRST current (27). In what follows, we shall comment on its nilpotency property.

The explicit form of the BRST charge $Q_{b}$, derived from (27), is as follows:
\begin{eqnarray}
&&Q_{b}  =\int d^{2} x\left[m \varepsilon^{0 i j} C_{i} \partial_{j} \tilde{\varphi}+\left(\partial^{0} \bar{C}^{i}-\partial^{i} \bar{C}^{0}\right) \partial_{i} \beta-\left(\partial^{0} C-m C^{0}\right) B\right.  \nonumber\\
&& -m\left(\partial^{0} \bar{C}-m \bar{C}^{0}\right) \beta+\left(m B^{0 i}-\Sigma^{0 i}\right)\left(\partial_{i} C-m C_{i}\right)-\rho \partial^{0} \beta-\lambda B^{0}  \nonumber\\
&& \left.-\frac{1}{2} \varepsilon^{0 i j} \mathcal{B}\left(\partial_{i} C_{j}-\partial_{j} C_{i}\right)-B^{i}\left(\partial_{0} C_{i}-\partial_{i} C_{0}\right)\right]. 
\end{eqnarray}
The above BRST charge is the generator for the BRST symmetry transformations in (26). For this purpose, we have to compute the conjugate momenta corresponding to {\it all} the basic fields of our theory and check that the following is true, namely;
\begin{eqnarray}
s_{b} \Phi(\vec{x}, t)=-i\left[\Phi(\vec{x}, t), Q_{b}\right]_{( \pm)}, 
\end{eqnarray}
where the $(\pm)$ signs, as the subscripts on the square bracket, stand for the bracket to be an (anti)commutator for the generic field $\Phi$ of our theory being fermionic/bosonic in nature. It is worthwhile to point out that the relationship between the continuous symmetry and the Noether conserved charge is sacrosanct and it is true for any kind of fermionic/bosonic field operator (chosen for the generic field $\Phi$). In this connection, it 
is very interesting to note that we have the sanctity of the following relationship
\begin{eqnarray}
s_{b} Q_{b}=-i\left\{Q_{b}, Q_{b}\right\} \equiv-2 i Q_{b}^{2}, 
\end{eqnarray}
that exist between the BRST symmetry transformation operator $s_{b}$ and the corresponding conserved Noether BRST charge $Q_{b}$. In the above, we have taken into account the fermionic nature of $Q_{b}$ which is evident from its explicit expression in (30). A close look at (32) shows that we can talk about the nilpotency property of the Noether BRST charge $Q_{b}$ if we can compute explicitly the l.h.s. of (32) by taking into account the BRST symmetry transformations (26) and the explicit expression for the BRST charge $Q_{b}$
[cf. Eq. (30)]. In fact, the explicit computation of the l.h.s. of (32) yields the following:
\begin{eqnarray}
s_{b} Q_{b}=\int d^{2} x\left[\left(\partial^{0} B^{i}-\partial^{i} B^{0}\right) \partial_{i} \beta-m\left(\partial^{0} B
-m B^{0}\right) \beta\right] \neq 0. 
\end{eqnarray}
A close look at (32) and (33) establishes the fact that the BRST charge $Q_{b}$ is not nilpotent (i.e. $Q_{b}^{2} \neq 0$) of order two. In the next subsection, we derive the nilpotent version where the emphasis is laid on the precise computation of the l.h.s. of (32) without spoiling the conservation law of the Noether BRST charge. \\

\subsection{Conserved BRST Charge: Nilpotent Version}

The central purpose of this subsection is to derive the nilpotent version of the BRST charge from the non-nilpotent Noether BRST charge. We have proposed a theoretical framework (see, e.g. [45] for details) which allows the derivation of the conserved and nilpotent version of the BRST charge $Q_{B}$ from the non-nilpotent Noether conserved charge $Q_{b}$. In fact, for this purpose, we have exploited the interplay of (i) the Gauss divergence theorem, (ii) the appropriate EL-EoMs, and (iii) the application of the nilpotent BRST transformations
at appropriate places. For instance, let us focus on the last but one term of the Noether conserved BRST charge $Q_{b}$ [cf. Eq. (30)] which can be re-written as
\begin{eqnarray}
-\int d^{2} x \varepsilon^{0 i j} \mathcal{B}\left(\partial_{i} C_{j}\right)=\int d^{2} x\left(\varepsilon^{0 i j} \partial_{i} \mathcal{B}\right) C_{j}, 
\end{eqnarray}
where we have exploited the theoretical potential of the Gauss divergence theorem. At this stage, we can use the last entry of the equation of motion (29) that has been derived from the BRST invariant Lagrangian density w.r.t. the massive gauge field $B_{\mu \nu}$. The following input coming out from the above equation of motion, namely;
\begin{eqnarray}
\left(\varepsilon^{0 i j} \partial_{i} \mathcal{B}\right) C_{j}=m\left(m B^{0 i}-\Sigma^{0 i}\right) C_{i}-m \varepsilon^{0 i j} C_{i} \partial_{j} \tilde{\varphi}+\left(\partial^{0} B^{i}-\partial^{i} B^{0}\right) C_{i},
\end{eqnarray}
leads to following form of the r.h.s. of (34), namely;
\begin{eqnarray}
\int d^{2} x\left[m\left(m B^{0 i}-\Sigma^{0 i}\right) C_{i}-m \varepsilon^{0 i j} C_{i} \partial_{j} \tilde{\varphi}+\left(\partial^{0} B^{i}-\partial^{i} B^{0}\right) C_{i}\right]. 
\end{eqnarray}
It is worthwhile to point out that, in the derivation of (35), we have used the explicit form of the last entry of (29) as: $\varepsilon^{\mu \nu \sigma} \partial_{\mu} \mathcal{B}+\left(\partial^{\nu} B^{\sigma}-\partial^{\sigma} B^{\nu}\right)=-m\left(m B^{\nu \sigma}-\Sigma^{\nu \sigma}\right)+m \varepsilon^{\nu \sigma \eta} \partial_{\eta} \tilde{\varphi}$ and taken the choice $\nu=0, \sigma=j$. We have also used the standard rules of the summation convention to express the r.h.s. of (35) in a compact and nice looking form. A close look at the r.h.s. of the above equation (36) and the expression for the Noether BRST charge [cf. Eq. (30)] demonstrates that the first two terms of the above equation will cancel out with (i) the first term of $Q_{b}$
[cf. Eq. (30)], and (ii) the second term of the r.h.s. of
\begin{eqnarray}
\left(m B^{0 i}-\Sigma^{0 i}\right)\left(\partial_{i} C-m C_{i}\right)=\left(m B^{0 i}-\Sigma^{0 i}\right) \partial_{i} C-m\left(m \,B^{0 i}-\Sigma^{0 i}\right) C_{i},
\end{eqnarray}
which is nothing but the expanded form the {\it fifth} term of the Noether conserved charge $Q_{b}$ [cf. Eq. (30)]. It is crystal clear, from the above exercise, that the first term, fifth term and the last but one term of the Noether conserved charge $Q_{b}$ [cf. Eq. (30)] lead to the existence of only the last term of (36) and the first term of (37) which will be present in the nilpotent (i.e. $Q_{B}^{2}=0$ ) version of $Q_{B}$ (which we derive from $Q_{b}$).

At this juncture, let us focus on the first term of (37) which is present inside the integration. Once again, we apply the Gauss divergence theorem to obtain the following:
\begin{eqnarray}
\int d^{2} x\left(m B^{0 i}-\Sigma^{0 i}\right) \partial_{i} C \equiv-\int d^{2} x \,\partial_{i}\left(m B^{0 i}-\Sigma^{0 i}\right) C.
\end{eqnarray}
We use, in (38), the first entry of the equations of motion in (29) to obtain the following
\begin{eqnarray}
\int d^{2} x\left(m\, B^{0}+\partial^{0} B\right) C, 
\end{eqnarray}
which will be present in the nilpotent version of the BRST charge $Q_{B}$. To sum up, we have two very important and useful terms of the nilpotent version of the BRST charge $Q_{B}$ as
\begin{eqnarray}
\int d^{2} x\left[\left(m B^{0}+\partial^{0} B\right) C+\left(\partial^{0} B^{i}-\partial^{i} B^{0}\right) C_{i}\right].
\end{eqnarray}
As per the rules illustrated in our earlier work [45], we have to apply the BRST symmetry transformations on the above two terms and re-arrange some of the relevant terms of $Q_{b}$ so that there are perfect cancellations. For instance, we note that the application of the BRST symmetry transformations
(26)  on the above two terms yield:
\begin{eqnarray}
-\int d^{2} x\left[\left(\partial^{0} B^{i}-\partial^{i} B^{0}\right) \partial_{i} \beta+m\left(m B^{0}+\partial^{0} B\right) \beta\right].
\end{eqnarray}
For the cancellation of the above two terms, we have to modify the second and fourth terms of the Noether BRST charge $Q_{b}[$ cf. Eq. (30)] as follows:
\begin{eqnarray}
&&\int d^{2} x\left(\partial^{0} \bar{C}^{i}-\partial^{i} \bar{C}^{0}\right) \partial_{i} \beta  =2 \int d^{2} x\left(\partial^{0} \bar{C}^{i}-\partial^{i} \bar{C}^{0}\right) \partial_{i} \beta-\int d^{2} x\left(\partial^{0} \bar{C}^{i}-\partial^{i} \bar{C}^{0}\right) \partial_{i} \beta,  \nonumber\\
&&-m \int d^{2} x\left(\partial^{0} \bar{C}-m \bar{C}^{0}\right) \beta =-2 m \int d^{2} x\left(\partial^{0} \bar{C}-m \bar{C}^{0}\right) \beta  \nonumber\\
&& +m \int d^{2} x\left(\partial^{0} \bar{C}-m \bar{C}^{0}\right) \beta.
\end{eqnarray}
It is straightforward to note that the application of the BRST symmetry transformations on the second entries on the r.h.s. of the above top as well as bottom equations leads to the perfect cancellation of (41). Thus, in addition to the two terms of (40), the above second entries on the r.h.s. of the top and bottom equations of (42) will also be present in the nilpotent version of the BRST charge $Q_{B}$. Let us now concentrate on the first entry that is present on the r.h.s. of the top equation of (42), namely;
\begin{eqnarray}
2 \int d^{2} x\left(\partial^{0} \bar{C}^{i}-\partial^{i} \bar{C}^{0}\right) \partial_{i} \beta \equiv+2 \int d^{2} x \partial_{i}\left(\partial^{i} \bar{C}^{0}-\partial^{0} \bar{C}^{i}\right) \beta,
\end{eqnarray}
where we have used the Gauss divergence theorem to throw away the total space derivative term. Using the equation of motion: $\partial_{\mu}\left(\partial^{\mu} \bar{C}^{\nu}-\partial^{\nu} \bar{C}^{\mu}\right)=\partial^{\nu} \rho+m\left(\partial^{\nu} \bar{C}-m \bar{C}^{\nu}\right)$ for the choice $\nu=0$, we obatin the following expression from the r.h.s. of (43), namely;
\begin{eqnarray}
+\int d^{2} x\left[2 \dot{\rho} \beta+2 m\left(\partial^{0} \bar{C}-m \bar{C}^{0}\right) \beta\right].
\end{eqnarray}
It can be readily checked that the second term on the r.h.s. of the above equation cancels out with the first term of the r.h.s. of (42). It is interesting to point out that the first term of the above equitation (44) is a BRST invariant quantity [i.e. $\left.s_{b}(2 \dot{\rho} \beta)=0\right]$. Thus, finally, the most important, useful and non-trivial five existing terms of the nilpotent (i.e. $\left.Q_{B}^{2}=0\right)$ version of the BRST charge $Q_{B}$ are as follows
\begin{eqnarray}
& \int d^{2} x\left[\left(m B^{0}+\partial^{0} B\right) C+\left(\partial^{0} B^{i}-\partial^{i} B^{0}\right) C_{i}\right.  \nonumber\\
& \left.+m\left(\partial^{0} \bar{C}-m \bar{C}^{0}\right) \beta-\left(\partial^{0} \bar{C}^{i}-\partial^{i} \bar{C}^{0}\right) \partial_{i} \beta+2 \dot{\rho} \beta\right], 
\end{eqnarray}
which are the sum of terms in (40), second terms of the r.h.s. of the bottom and top equations of (42) and the first term of (44). It is crystal clear that if we apply the BRST symmetry transformations $s_{b}$ on the above terms, it turns out to be zero.

We have obtained the non-trivial and very useful BRST invariant {\it five} terms of (45) by focusing on the first, second, fourth, fifth and eighth terms of the total nine terms that are present in the conserved (but non-nilpotent) Noether conserved charge $Q_{b}$ [cf. Eq. (30)]. It is interesting to point out that all the rest of the terms in $Q_{b}$ are BRST invariant quantities. Hence, finally, the explicit expression for the nilpotent (i.e. $Q_{B}^{2}=0$ ) version of the BRST charge $Q_{B}$ (derived from the non-nilpotent Noether conserved charge $Q_{b}$) is:
\begin{eqnarray}
&& Q_{b} \rightarrow Q_{B}=\int d^{2} x\left[\left(\partial^{0} B^{i}-\partial^{i} B^{0}\right) C_{i}+\left(m B^{0}+\partial^{0} B\right) C+m\left(\partial^{0} \bar{C}-m \bar{C}^{0}  \right) \beta  -\lambda B^{0} \right. \nonumber\\
&& \left.-\left(\partial^{0} \bar{C}^{i}-\partial^{i} \bar{C}^{0}\right) \partial_{i} \beta+2 \dot{\rho} \beta-\rho \dot{\beta}-B^{i}\left(\partial_{0} C_{i}-\partial_{i} C_{0}\right)-\left(\partial^{0} C-m C^{0}\right) B\right].
\end{eqnarray}
It is straightforward to note that if we apply the BRST symmetry transformations $s_{b}$ on the above expression, we obtain zero. In other words, we find that: $s_{b} Q_{B}=-i\left\{Q_{B}, Q_{B}\right\}=0$. This implies that the BRST charge $Q_{B}$ (that has been systematically derived from the non-nilpotent Noether conserved charge $Q_{b}$) is indeed nilpotent (i.e. $Q_{B}^{2}=0$) of order two.

We end this subsection with the following remarks. First of all, the existence of the conserved (but non-nilpotent) version of the Noether charge is a clear indication that there are non-trivial CF-type restrictions on our theory (see, e.g. [45] for details). Second, to obtain the nilpotent version of the BRST charge $Q_{B}$ (from the non-nilpotent version of the Noether charge $Q_{b}$), we have exploited only the appropriate EL-EoMs and Gauss's divergence theorem. Thus, the nilpotent version of the BRST charge $Q_{B}$ is conserved (just like the Noether conserved charge $Q_{b}$). Third, we know that our BRST-quantized theory is endowed with the non-trivial CF-type restrictions. We shall derive these very important (anti-)BRST invariant CF-type restrictions, from different theoretical angles, in our Sec. 5 (see below). Finally, the nilpotency property is very important because (i) the BRST cohomology crucially depends on this property, and (ii) the operator forms of the first-class constraints of our theory appear in the physicality criteria (i.e. $Q_{(A) B}\,|phys >=0$) w.r.t. only the nilpotent (i.e. $Q_{(A) B}^{2}=0$) versions of the (anti-)BRST charges $Q_{(A) B}$ where they annihilate the physical states (i.e. $|phys >$). The latter observation is important because it is consistent with the Dirac quantization conditions for systems that are endowed with any kinds of constraints (in the terminology of Dirac's classification scheme).\\

\section{Anti-BRST Transformations: Lagrangian Density and Nilpotency Property of the Anti-BRST Charge}

Our present section contains two subsections. In Subsec. 4.1, we derive the conserved Noether anti-BRST current and corresponding charge $Q_{a b}$ and show that the latter is nonnilpotent (i.e. $Q_{a b}^{2} \neq 0$) of order two. This happens due to the existence of the {\it non-trivial} CF-type restrictions on our theory
(see, e.g. [45] for details).
We briefly comment that the Noether conserved anti-BRST charge $Q_{a b}$ is the generator for the infinitesimal, continuous and off-shell nilpotent anti-BRST symmetry transformations [cf. Eq. (49) below]. Our Subsec. 4.2 deals with the derivation of the nilpotent (i.e. $Q_{A B}^{2}=0$) version of the anti-BRST charge $Q_{A B}$ from its non-nilpotent counterpart Noether anti-BRST charge $Q_{a b}$. We also provide a few simple arguments to establish that the nilpotent version of the anti-BRST charge $Q_{A B}$ is {\it also} conserved (just like the Noether anti-BRST charge $Q_{a b}$).\\

\subsection{Anti-BRST Symmetries: Noether Conserved Charge}

Against the backdrop of our elaborate discussions in the previous section, we shall be brief in our present subsection where we start with the analogue of the gauge-fixed Lagrangian density (22) for our modified massive 3D Abelian 2-form theory as follows
\begin{eqnarray}
\mathcal{L}_{(0)}^{(M S)}+\mathcal{L}_{(g f)}^{(B)} & =&\frac{1}{2}\left(H_{012}+m \tilde{\varphi}\right)^{2}-\frac{m^{2}}{4} B^{\mu \nu} B_{\mu \nu}+\frac{m}{2} B^{\mu \nu}\left[\Sigma_{\mu \nu}+\varepsilon_{\mu \nu \sigma} \partial^{\sigma} \tilde{\varphi}\right]-\frac{1}{4} \Sigma^{\mu \nu} \Sigma_{\mu \nu} \nonumber\\
& -&\frac{1}{2} \partial_{\mu} \tilde{\varphi} \partial^{\mu} \tilde{\varphi}-\frac{1}{2}(\partial \cdot \phi-m \phi)^{2}+\frac{1}{2}\left(\partial^{\nu} B_{\nu \mu}+\partial_{\mu} \phi+m \phi_{\mu}\right)^{2},
\end{eqnarray}
where it is worthwhile to mention that (i) we have taken into account $\phi \rightarrow-\phi$ for the scalar field so that it differs from the gauge-fixed-Lagrangian density (22) and provides more generality to the 't Hooft gauge, and (ii) we still have the validity of the on-shell condition $\left(\square+m^{2}\right) \phi=0$ that is also true for the Lagrangian density (22). As we have pointed out in our earlier 
works [14,15,,25,26], it is very interesting to mention that the scalar and pseudo-scalar fields of our theory satisfy the Klein-Gordon equations despite the fact that they carry the kinetic terms that are endowed with opposite signs. The above quadratic gauge-fixed Lagrangian density (with the quadratic kinetic term for the massive gauge field $B_{\mu\nu}$) can be linearized by invoking another set of {\it three} 
Nakanishi-Lautrup auxiliary fields (e.g. $\mathcal{B}, \bar{B}, \bar{B}_{\mu}$ ) as:
\begin{eqnarray}
&& \mathcal{L}_{(\bar{b})}=\mathcal{B}\left(\frac{1}{2} \varepsilon^{\mu \nu \sigma} \partial_{\mu} B_{\nu \sigma}+m \tilde{\varphi}\right)-\frac{\mathcal{B}^{2}}{2}-\frac{m^{2}}{4} B^{\mu \nu} B_{\mu \nu}+\frac{m}{2} B^{\mu \nu}\left(\Sigma_{\mu \nu}+\varepsilon_{\mu \nu \sigma} \partial^{\sigma} \tilde{\varphi}\right)+\frac{\bar{B}^{2}}{2} \nonumber\\
&& +\bar{B}(\partial \cdot \phi-m \phi)-\frac{1}{4} \Sigma^{\mu \nu} \Sigma_{\mu \nu}-\frac{1}{2} \partial_{\mu} \tilde{\varphi} \partial^{\mu} \tilde{\varphi}+\bar{B}^{\mu}\left(\partial^{\nu} B_{\nu \mu}+\partial_{\mu} \phi+m \phi_{\mu}\right)-\frac{\bar{B}^{\mu} \overline{\mathcal{B}}_{\mu}}{2}.
\end{eqnarray}
It will be worthwhile to mention here that the quadratic kinetic term for the massive gauge field [i.e. the first term in (47)] remains the same as nothing can be altered in it. Hence, the auxiliary field $\mathcal{B}$ remains the same as in (23). The subscript $(\bar{b})$ on the above Lagrangian density denotes that it is different from (23) and is meant for our discussion on the nilpotent anti-BRST symmetries where the FP-ghost terms remain the same as in (24). Thus, the total anti-BRST invariant Lagrangian density $\mathcal{L}_{B}$ is the sum of (48) and (24) 
[i.e. $\left.\mathcal{L}_{\bar{B}}=\mathcal{L}_{(\bar{b})}+\mathcal{L}_{(F P)}\right]$. Under the following infinitesimal, continuous and off-shell nilpotent (i.e. $s_{a b}^{2}=0$ ) anti-BRST symmetry transformations $s_{a b}$
\begin{eqnarray}
&& s_{a b} B_{\mu \nu}=-\left(\partial_{\mu} \bar{C}_{\nu}-\partial_{\nu} \bar{C}_{\mu}\right), \quad s_{a b} \bar{C}_{\mu}=-\partial_{\mu} \bar{\beta}, \quad s_{a b} C_{\mu}=\bar{B}_{\mu}, \nonumber\\
&& s_{a b} \phi_{\mu}=\left(\partial_{\mu} \bar{C}-m \bar{C}_{\mu}\right), \quad s_{a b} \bar{C}=-m \bar{\beta}, \quad s_{a b} C=\bar{B}, \nonumber\\
&& s_{a b} \beta=-\lambda, \quad s_{a b} \phi = +\rho, \quad s_{a b} \Sigma_{\mu \nu}=-m\left(\partial_{\mu} \bar{C}_{\nu}-\partial_{\nu} \bar{C}_{\mu}\right), \nonumber\\
&& s_{a b}\left[\tilde{\varphi}, \rho, \lambda, \bar{\beta}, \bar{B}_{\mu}, \bar{B}, \mathcal{B}, H_{\mu \nu \lambda}\right]=0,
\end{eqnarray}
the total Lagrangian density $\left(\mathcal{L}_{\bar{B}}\right)$ transforms to the total spacetime derivative as
\begin{eqnarray}
s_{a b} \mathcal{L}_{\bar{B}}=\partial_{\mu}\left\{\rho \bar{B}^{\mu}-\lambda \partial^{\mu} \bar{\beta}+\left(\partial^{\mu} \bar{C}-m \bar{C}^{\mu}\right) \bar{B}-m \varepsilon^{\mu \nu \sigma} \bar{C}_{\nu} \partial_{\sigma} \tilde{\varphi}-\left(\partial^{\mu} \bar{C}^{\nu}-\partial^{\nu} \bar{C}^{\mu}\right) \bar{B}_{\nu}\right\},
\end{eqnarray}
which establishes that the action integral $S=\int d^{3} x \mathcal{L}_{\bar{B}}$ remains invariant (i.e. $s_{a b} S=0$ ) under the continuous, off-shell nilpotent and infinitesimal anti-BRST symmetry transformations (49) because the physically well-defined fields vanish off as $x \rightarrow \pm \infty$.

According to Noether's theorem, the invariance of the action integral under a set of infinitesimal continuous symmetry transformations leads to the derivation of the conserved Noether current. Following the analogue of (5), we obtain the following explicit expression for the Noether anti-BRST current 
($J_{(ab)}^{\mu}$):
\begin{eqnarray}
J_{(a b)}^{\mu} & =&m \varepsilon^{\mu \nu \sigma} \bar{C}_{\nu} \partial_{\sigma} \tilde{\varphi}+m\left(\partial^{\mu} C-m C^{\mu}\right) \bar{\beta}-\lambda \partial^{\mu} \bar{\beta}-\left(\partial^{\mu} C^{\nu}-\partial^{\nu} C^{\mu}\right) \partial_{\nu} \bar{\beta} 
\nonumber\\
& +&\left(m B^{\mu \nu}-\Sigma^{\mu \nu}\right)\left(\partial_{\nu} \bar{C}-m \bar{C}_{\nu}\right)+\rho \bar{B}^{\mu}+\left(\partial^{\mu} \bar{C}-m \bar{C}^{\mu}\right) \bar{B} \nonumber\\
& -&\frac{1}{2}\left[\varepsilon^{\mu \nu \sigma} \mathcal{B}+\left(\eta^{\mu \nu} \bar{B}^{\sigma}-\eta^{\mu \sigma} \bar{B}^{\nu}\right)\right]\left(\partial_{\nu} \bar{C}_{\sigma}-\partial_{\sigma} \bar{C}_{\nu}\right).
\end{eqnarray}
The above current is conserved (i.e. $\partial_{\mu}  J_{(a b)}^{\mu}=0$) provided we use the equations of motion (28) that emerge out from the ghost-sector [i.e. $\mathcal{L}_{(F P)}$] of the total Lagrangian density and the following EL-EoMs that are derived from the non-ghost sector [i.e. $\mathcal{L}_{(\bar{b})}$ ], namely;
\begin{eqnarray}
&& \varepsilon^{\mu \nu \sigma} \partial_{\mu} \mathcal{B}+\left(\partial^{\nu} \bar{B}^{\sigma}-\partial^{\sigma} \bar{B}^{\nu}\right)=-m\left(m B^{\nu \sigma}-\Sigma^{\nu \sigma}\right)+m \varepsilon^{\nu \sigma \eta} \partial_{\eta} \tilde{\varphi}, \nonumber\\
&& \partial_{\mu}\left(m B^{\mu \nu}-\Sigma^{\mu \nu}\right)=m \bar{B}^{\nu}-\partial^{\nu} \bar{B}, \quad \partial_{\mu} \bar{B}^{\mu}=-m \bar{B}.
\end{eqnarray}
w.r.t. the massive antisymmetric (i.e. $B_{\mu \nu}=-B_{\nu \mu}$) tensor gauge field $B_{\mu \nu}$, vector field $\phi_{\mu}$ and the scalar field $\phi$, respectively. From the conserved Noether anti-BRST current (51), we can derive the conserved anti-BRST charge (i.c. $Q_{a b}=\int d^{2} x J_{(a b)}^{0}$) as follows:
\begin{eqnarray}
&&Q_{a b}  =\int d^{2} x\left[\rho \bar{B}^{0}-\lambda \dot{\bar{\beta}}+\left(\partial^{0} \bar{C}-m \bar{C}^{0}\right) \bar{B}+m\left(\partial^{0} C-m C^{0}\right) \bar{\beta}+m \varepsilon^{0 i j} \bar{C}_{i} \partial_{j} \tilde{\varphi}\right. \nonumber\\
&& +\left(m B^{0 i}-\Sigma^{0 i}\right)\left(\partial_{i} \bar{C}-m \bar{C}_{i}\right)-\left(\partial^{0} C^{i}-\partial^{i} \bar{C}^{0}\right) \partial_{i} \bar{\beta}-\frac{1}{2} \varepsilon^{o i j} \mathcal{B}\left(\partial_{i} \bar{C}_{j}-\partial_{j} \bar{C}_{i}\right) \nonumber\\
&& \left.-\bar{B}^{i}\left(\partial_{0} \bar{C}_{i}-\partial_{i} \bar{C}_{0}\right)\right]. 
\end{eqnarray}
The above charge is the generator for the infinitesimal, continuous and off-shell nilpotent anti-BRST symmetry transformations (49) provided we compute the conjugate momentum corresponding to the generic field $\Phi$ of our theory and use the equation (31) with the replacements: 
$s_{b} \rightarrow s_{ab}, \;Q_{b} \rightarrow Q_{a b}$. However, it is very interesting to point out that the above Noether conserved anti-BRST charge $Q_{a b}$ is found to be non-nilpotent (i.c. $Q_{a b}^{2} \neq 0$) of order two because we observe that the following is true, namely;
\begin{eqnarray}
s_{a b} Q_{a b}=-i\left\{Q_{a b}, Q_{a b}\right\} \equiv-2 i Q_{a b}^{2} \neq 0.
\end{eqnarray}
To corroborate the above statement, we have to compute explicitly the l.h.s. of (54) by using (49) and the explicit expression for $Q_{a b}$
[cf. Eq. (53)]. To be precise, we note that the following is true as far as the computation of the l.h.s. of (54) is concerned, namely;
\begin{eqnarray}
s_{a b} Q_{a b}=\int d^{2} x\left[m\left(\partial^{0} \bar{B}-m \bar{B}^{0}\right) \bar{\beta}-\left(\partial^{0} \bar{B}^{i}-\partial^{i} \bar{B}^{0}\right) \partial_{i} \bar{\beta}\right] \neq 0.
\end{eqnarray}
In other words, we find that the Noether anti-BRST charge is conserved and it is the generator for the infinitesimal, continuous and off-shell nilpotent anti-BRST symmetry transformations (49). However, it is not nilpotent of order two (i.e. $\left.Q_{a b}^{2} \neq 0\right)$.\\

\subsection{Conserved Anti-BRST Charge: Nilpotent Version}

With the background of our elaborate discussions in Subsec. 3.2, we shall be brief in our present subsection where we shall derive the 
nilpotent (i.e. $Q_{A B}^{2}=0$) version of the anti-BRST charge $Q_{A B}$ from the Noether anti-BRST charge $Q_{a b}$ [cf. Eq. (53)]. In this context, first of all, we focus on the last but one term of (53) which can be re-expressed a
\begin{eqnarray}
-\int d^{2} x \varepsilon^{0 i j} \mathcal{B}\left(\partial_{i} \bar{C}_{j}\right)=\int d^{2} x\left(\varepsilon^{0 i j} \partial_{i} \mathcal{B}\right) \bar{C}_{j},
\end{eqnarray}
where we have dropped a total space derivative term due to Gauss's divergence theorem. Using the EL-EoMs (52), the r.h.s. of the above equation can be re-written as:
\begin{eqnarray}
\int d^{2} x\left[m\left(m B^{0 i}-\Sigma^{0 i}\right) \bar{C}_{i}-m \varepsilon^{0 i j} \bar{C}_{i} \partial_{j} \tilde{\varphi}+\left(\partial^{0} B^{i}-\partial^{i} B^{0}\right) \bar{C}_{i}\right]. 
\end{eqnarray}
To be precise, we have exploited the beauty of the top entry in (52) which is nothing but the EL-EoM w.r.t. the massive gauge field $B_{\mu \nu}$. Furthermore, the {\it sixth} term of the Noether anti-BRST charge (53) can be expanded as:
\begin{eqnarray}
\left(m B^{0 i}-\Sigma^{0 i}\right)\left(\partial_{i} \bar{C}-m \bar{C}_{i}\right)=\left(m B^{0 i}-\Sigma^{0 i}\right) \partial_{i} \bar{C}-m\left(m B^{0 i}-\Sigma^{0 i}\right) \bar{C}_{i}. 
\end{eqnarray}
A close and careful look at the equations (53), (57) and (58) demonstrate that the first two terms of (57) would cancel out with the 
{\it second} term of (58) and the {\it fifth} term  of (53). Hence, only the last term will survive in (57) and it will be present in the nilpotent version (i.e. $Q_{A B}^{2}=0$) of the anti-BRST charge $Q_{A B}$. Now we concentrate on the first term of (58) which is present inside the integral. Using the Gauss divergence theorem, it can be re-written and re-expressed, using the appropriate EL-EoM from (52), as:
\begin{eqnarray}
-\int d^{2} x \partial_{i}\left(m B^{0 i}-\Sigma^{0 i}\right) \bar{C}=
+\int d^{2} x\left(m \bar{B}^{0}-\partial^{0} \bar{B}\right) \bar{C}.
\end{eqnarray}
This term will {\it also} be a part of the nilpotent version (i.e. $Q_{A B}^{2}=0$) of the anti-BRST charge $Q_{A B}$. Thus, the last term of (57) and the above term are a part of $Q_{A B}$ which can be re-expressed in the following manner:
\begin{eqnarray}
\int d^{2} x\left[\left(\partial^{0} \bar{B}^{i}-\partial^{i} \bar{B}^{0}\right) \bar{C}_{i}-\left(\partial^{0} \bar{B}-m \bar{B}^{0}\right) \bar{C}\right].
\end{eqnarray}
At this juncture, we apply the anti-BRST symmetry transformations (49) on the above expression and re-arrange some of the appropriate terms of the Noether anti-BRST charge (53) so that there are perfect cancellations. For instance, the application of the anti-BRST symmetry transformations on the above terms of (60) leads to the following:
\begin{eqnarray}
\int d^{2} x\left[m\left(\partial^{0} \bar{B}-m \bar{B}^{0}\right) \bar{\beta}-\left(\partial^{0} \bar{B}^{i}-\partial^{i} \bar{B}^{0}\right) \partial_{i} \bar{\beta}\right].
\end{eqnarray}
Now the stage is suitable set to modify some of the appropriate terms of (53) so that when the anti-BRST symmetry transformations are applied on a part of them, there are perfect cancellations between them and (61). Towards this goal in mind, we re-express a couple of appropriate terms of (53) as follows
\begin{eqnarray}
&&m \int d^{2} x\left(\partial^{0} C-m C^{0}\right) \bar{\beta}  =2 m \int d^{2} x\left(\partial^{0} C-m C^{0}\right) \bar{\beta}-m \int d^{2} x\left(\partial^{0} C-m C^{0}\right) \bar{\beta}, \nonumber\\
&&-\int d^{2} x\left(\partial^{0} C^{i}-\partial^{i} \bar{C}^{0}\right) \partial_{i} \bar{\beta}  =-2 \int d^{2} x\left(\partial^{0} C^{i}-\partial^{i} \bar{C}^{0}\right) \partial_{i} \bar{\beta} \nonumber\\
&& +\int d^{2} x\left(\partial^{0} C^{i}-\partial^{i} \bar{C}^{0}\right) \partial_{i} \bar{\beta},
\end{eqnarray}
which are nothing but the {\it third} and {\it seventh}
 terms of the Noether charge $Q_{a b}$. It is straightforward to note that if apply the anti-BRST symmetry transformations (49) on the second terms on the r.h.s. of the top and and bottom equations, the resulting expressions cancel out with both the terms of (61). Thus, in addition to (61), the second terms on the r.h.s. of the above equation ( 62 ) would also be present in $Q_{A B}$.

To obtain the final form of the nilpotent (i.e. $Q_{A B}^{2}=0$) version of the anti-BRST charge $Q_{A B}$, we concentrate on the first integral on the r.h.s. of the bottom equation of (62). The latter can be re-expressed, using the Gauss divergence theorem, as
\begin{eqnarray}
+2 \int d^{2} x \partial_{i}\left(\partial^{0} C^{i}-\partial^{i} \bar{C}^{0}\right) \bar{\beta} \equiv-2 \int d^{2} x \partial_{i}\left(\partial^{i} C^{0}-\partial^{0} C^{i}\right) \bar{\beta}. 
\end{eqnarray}
It is obvious that if we use the EL-EoM: $\partial_{\mu}\left(\partial^{\mu} C^{\nu}-\partial^{\nu} C^{\mu}\right)=-\partial^{\nu} \lambda+m\left(\partial^{\nu} C-m C^{\nu}\right)$ for the choice: $\nu=0$, the r.h.s. of the above equation (63) can be re-written as
\begin{eqnarray}
\int d^{2} x\left[+2 \dot{\lambda} \bar{\beta}-2 m\left(\partial^{0} C-m C^{0}\right) \bar{\beta}\right].
\end{eqnarray}
A straightforward observation of (62) and (64) shows that the last entry of the {\it latter} will cancel out with the first entry on the r.h.s. of the top equation of the former. Thus, ultimately, we have obtained five crucial terms of $Q_{A B}$ as follows
\begin{eqnarray}
& \int d^{2} x\left[\left(\partial^{0} \bar{B}^{i}-\partial^{i} \bar{B}^{0}\right) \bar{C}_{i}-\left(\partial^{0} \bar{B}-m \bar{B}^{0}\right) \bar{C}-m\left(\partial^{0} C-m C^{0}\right) \bar{\beta}\right.\nonumber\\
& \left.+\left(\partial^{0} C^{i}-\partial^{i} \bar{C}^{0}\right) \partial_{i} \bar{\beta}+2 \dot{\lambda} \bar{\beta}\right],
\end{eqnarray}
which are nothing but the sum of (61), second terms on the r.h.s. of top and bottom equations of (62) and the first entry in (65). It is pretty clear that the latter entry is an anti-BRST invariant [i.e. $s_{a b}(2 \dot{\lambda} \bar{\beta})=0$] quantity. Thus, we have taken into account all the non-trivial terms of the Noether conserved (but non-nilpotent) charge $Q_{a b}$. The rest of the terms of $Q_{a b}$ are trivially anti-BRST invariant. Hence, the final form of the conserved and nilpotent version of the anti-BRST charge $Q_{A B}$ is as follows:
\begin{eqnarray}
& Q_{ab} \to Q_{A B}=\int d^{2} x\left[\left(\partial^{0} \bar{B}^{i}-\partial^{i} \bar{B}^{0}\right) \bar{C}^{i}-\left(\partial^{0} \bar{B}-m \bar{B}^{0}\right) \bar{C}-m\left(\partial^{0} C-m C^{0}\right) \bar{\beta}\right. \nonumber\\
& \left.+\left(\partial^{0} C^{i}-\partial^{i} C^{0}\right) \partial_{i} \bar{\beta}+2 \dot{\lambda} \bar{\beta}-\lambda \dot{\bar{\beta}}+\rho \bar{B}^{0}+\left(\partial^{0} \bar{C}- m \bar{C}^{0}\right) \bar{B}-\bar{B}^{i}\left(\partial_{0} \bar{C}_{i}-\partial_{i} \bar{C}_{0}\right)\right].
\end{eqnarray}
The above is nothing but the sum of (65) and the rest of the BRST-invariant terms of $Q_{a b}\left[c f\right.$. Eq. (53)]. It is straightforward to check that we have the validity of $s_{a b} Q_{A B}=$ $-i\left\{Q_{A B}, Q_{A B}\right\}=0 \Rightarrow Q_{A B}^{2}=0$ which proves the off-shell nilpotency of $Q_{A B}$.

We wrap-up this subsection with the following two crucial remarks. First of all, as is well-known from Noether's theorem, the expression for the Noether conserved charge can be recast into any other form by exploiting the beauty and strength of the EL-EoMs for a given theory. This is what exactly we have followed in the derivation of $Q_{A B}$ from $Q_{a b}$ along with the standard definition of the physical fields that vanish off as $x \rightarrow \pm \infty$. The 
{\it latter} requirement has been taken care of by Gauss's divergence theorem which we have utilized at various places. Hence, if $Q_{a b}$ is conserved, the alternative expression $Q_{A B}$ (that is derived by using (i) the EL-EoMS, and (ii) the Gauss divergence theorem) will also be conserved. Second, the nilpotency property of the (anti-)BRST charges is very essential from the point of view of (i) the standard BRST algebra (cf. Appendix A below), (ii) the BRST cohomology (see, e.g. [42]), and (iii) the Dirac quantization conditions where the operator form of the constraints of the classical theory must annihilate the physical states (of the total quantum Hilbert space of states) at the quantum level (cf. Appendix B below). \\

\section{Curci-Ferrari Type Restrictions: Importance}

The existence of the (non-)trivial CF-type restriction(s) on a BRST-quantized theory is as fundamental as the existence of the first-class constraints for the definition of a classical gauge theory. In the case of our present theory, the (anti-)BRST invariant CF-type restrictions are responsible for (i) the absolute anticommutativity property 
of the BRST and anti-BRST symmetry transformations which encodes the linear independence of the BRST and anti-BRST symmetries, and (ii) the existence of the coupled (but equivalent) (anti-)BRST invariant Lagrangian densities which respect both BRST and anti-BRST symmetry transformations (provided we take into account the validity of the CF-type restrictions). Our present section contains three parts. In Subsec. 5.1, we derive the CF-type restrictions by demanding the direct equality of the BRST and anti-BRST invariant Lagrangian densities. Our Subsec. 5.2 deals with the proof that the (anti-)BRST invariant Lagrangian densities are {\it equivalent} w.r.t. the BRST and anti-BRST symmetry transformations in the sense that both the Lagrangian densities respect both the nilpotent symmetries (provided the validity of CF-type restrictions is invoked). Finally, in our Sebsec. 5.3, we establish the importance of the CF-type restrictions in the proof of the absolute anticommutativity between the BRST and anti-BRST symmetry transformations.\\

\subsection{Direct Equality of the Lagrangian Densities}

As pointed out earlier, the CF -type restrictions are responsible for the equivalence of the (anti-)BRST invariant Lagrangian densities modulo some total spacetime derivatives. The latter, as we know, do not play any significant role in the description of the dynamics of the theory within the framework of the Lagrangian formulation. To be precise, we wish to prove that the difference between the BRST invariant Lagrangian density $\mathcal{L}_{B}$ and anti-BRST invariant Lagrangian density $\mathcal{L}_{\bar{B}}$ is equal to the total spacetime derivative provided
we take into account the validity of the CF-type restrictions. We know that the FP-ghost part of the Lagrangian density (24) is common in the Lagrangian densities $\mathcal{L}_{B}$ and $\mathcal{L}_{\bar{B}}$. Hence, in the difference $\mathcal{L}_{B}-\mathcal{L}_{\bar B}$, it will cancel out. Ultimately, this 
difference reduces to $\mathcal{L}_{(b)}-\mathcal{L}_{(\bar b)}$ [cf. Eqs. (23),(48)] which leads to the following expression, namely;
\begin{eqnarray}
&& \frac{1}{2}(B+\bar{B})(B-\bar{B})-(B+\bar{B})(\partial \cdot \phi)-m(B-\bar{B})\, \phi-\frac{1}{2}\left(B^{\mu}-\bar{B}^{\mu}\right)\left(B_{\mu}+\bar{B}_{\mu}\right) \nonumber\\
&& +m\left(B^{\mu}-\bar{B}^{\mu}\right) \phi_{\mu}-\left(B^{\mu}+\bar{B}^{\mu}\right) \partial_{\mu} \phi+\left(B^{\mu}-\bar{B}^{\mu}\right)\left(\partial^{\nu} B_{\nu \mu}\right),
\end{eqnarray}
where we have used the simple algebraic trick of factorization of the difference between two square terms [e.g. $B^{2}-\bar{B}^{2}=(B+\bar{B})(B-\bar{B})$, etc.]. Re-arrangement of the above equation leads to the following form of the ensuing equation:
\begin{eqnarray}
&& \left(B^{\mu}-\bar{B}^{\mu}+ 2\, \partial^{\mu} \phi\right)
\left [\left(\partial^{\nu} B_{\nu \mu}\right)+m \phi_{\mu}-\frac{1}{2}\left(B_{\mu}+\bar{B}_{\mu}\right) \right] \nonumber\\
&& +(B+\bar{B}- 2\, m\, \phi) \left[\frac{1}{2}(B-\bar{B})-(\partial \cdot \phi)\right]+\partial_{\mu} \left[- \,2\, m\, \phi^{\mu}\, \phi \right ].
\end{eqnarray}
It is pretty obvious that if we impose the restrictions: $B+\bar{B}-2 m \phi=0$ and $B_{\mu}-\bar{B}_{\mu}+$ $2 \partial_{\mu} \phi=0$ on our theory, we observe that $\mathcal{L}_{B}$ and $\mathcal{L}_{\bar{B}}$ differ from 
each-other by a total spacetime derivative term: $\partial_{\mu}\left[-2 m \phi^{\mu} \phi\right]$. Hence, they are equivalent. The above two relationships among the Nakanishi-Lautrup auxiliary fields (e.g. $B, \bar{B}, B_{\mu}, \bar{B}_{\mu}$) and the scalar field $\phi$ are known as the CF-type restrictions which are, as is clear, responsible for the {\it equivalence} of the BRST and anti-BRST invariant Lagrangian densities.

The CF-type restrictions exist at the quantum level within the framework of BRST formalism. They are physical restrictions on our theory and, therefore, they should be (anti-)BRST invariant quantities. This requirement leads to the following off-shell nilpotent (anti-)BRST symmetry transformations for the Nakanishi-Lautrup auxiliary fields
\begin{eqnarray}\label{69}
s_{a b} B_{\mu}=- 2\, \partial_{\mu} \rho, \qquad s_{a b} B= 2\, m \,\rho, 
\qquad s_{b} \bar{B}_{\mu} = 2 \,\partial_{\mu} \lambda, \qquad s_{b} \bar{B}= 2\, m\, \lambda,
\end{eqnarray}
which are {\it not} present in the (anti-)BRST symmetry transformations that have been listed in (49) and (26), respectively. As a side remark, we would like
to mention that the above physically important and 
(anti-)BRST invariant non-trivial CF-type restrictions (i.e. $B_{\mu}-\bar{B}_{\mu}+ 2\,\partial_{\mu}\phi =0,\; B+\bar{B}- 2\,m \,\phi =0 $)
can emerge out from the following EL-EoMs
\begin{eqnarray}\label{70}
&& B_{\mu}=\partial^{\nu}B_{\nu \mu}-\partial_{\mu}\phi + m \, \phi_{\mu},   \qquad B =(\partial \cdot \phi)+ m\, \phi,  \nonumber\\
&& \bar{B}_{\mu}=\partial^{\nu}B_{\nu \mu}+\partial_{\mu}\phi + m \,\phi_{\mu}, \qquad \bar{B}=-(\partial \cdot \phi)+ m\,\phi,
\end{eqnarray}
which are derived from the Lagrangian densities ${\cal L}_B $ and 
${\cal L}_{\bar B} $ w.r.t. the Nakanishi-Lautrup auxiliary fields ($ B_\mu, B, \,\bar B_\mu,\, \bar B$). Hence, there is consistency between 
our observations in (68) and (70) as far as the validity of the CF-type restrictions are concerned.

We wrap-up this subsection with the remarks that (i) the {\it (non-)trivial} CF-type restrictions are an inevitable part of a properly BRST-quantized theory
as they are physical in the sense that they are (anti-)BRST invariant, (ii) the direct equality of the BRST and anti-BRST invariant 
coupled (but equivalent) Lagrangian densities leads to their derivation, and (iii) the EL-EoMs from these (anti-)BRST invariant Lagrangian densities
{\it also} give a glimpse of their existence on a properly BRST-quantized theory. However, we would like to add that the {\it latter} 
is an observation and it is {\it not} a systematic theoretical proof.\\

\subsection{Equivalent Lagrangian Densities: Symmetry Considerations}

We have already seen that, under the BRST symmetry transformations (26), the Lagrangian density ${\cal L}_B$ transforms to the total spacetime derivative
[cf. Eq. (25)]. In exactly similar fashion, we have noted that  ${\cal L}_{\bar B}$  transforms to the total spacetime derivative [cf. Eq. (50)] under
the anti-BRST symmetry transformations (49). The purpose of this subsection is to show that (i) the perfectly BRST invariant Lagrangian density 
${\cal L}_B$ also respects the anti-BRST symmetry transformations (49), and (ii) the perfectly anti-BRST invariant Lagrangian density 
${\cal L}_{\bar B}$ also respects the BRST symmetry transformations (26), provided we take into account the validity of the CF-type restrictions
(i.e. $B + \bar B - 2\, m \,\phi = 0, \; B_\mu- \bar B_\mu + 2 \, \partial_\mu\phi = 0$). In other words, when we apply the BRST symmetry transformations
($s_b$) on ${\cal L}_{\bar B}$ and anti-BRST symmetry transformations ($s_{ab}$) on ${\cal L}_B$, we obtain the total spacetime derivative terms
plus the terms that contain the above CF-type restrictions. To corroborate this statement, first of all, we focus on the computation of 
$s_b {\cal L}_{\bar B}$ which leads to the following explicit expression
\begin{eqnarray}
s_{b}\mathcal{L}_{\bar{B}} &=& \partial_{\mu}\Big[2(\partial_{\nu}B^{\nu \mu})\lambda -(\partial^{\mu}C^{\nu}-\partial^{\nu}C^{\mu})\bar{B}_{\nu}+(\partial^{\mu}C-mC^{\mu})\bar{B}-\rho \partial^{\mu}\beta +2m\lambda \phi^{\mu} \nonumber\\ 
&-&\lambda B^{\mu} -m\varepsilon^{\mu \nu \sigma}C_{\nu}\partial_{\sigma}\tilde{\varphi} \Big]-(\partial^{\mu}C^{\nu}-\partial^{\nu}C^{\mu})\partial_{\mu}\Big[B_{\nu}-\bar{B}_{\nu}+2\partial_{\nu}\phi\Big] \nonumber\\ 
&-&m(\partial^{\mu}C-mC^{\mu})(B_{\mu}-\bar{B}_{\mu}+2\partial_{\mu}\phi)+m\lambda(B+\bar{B}-2m\phi) \nonumber\\ 
&-&(\partial^{\mu}C-mC^{\mu})\partial_{\mu}\Big[B+\bar{B}-2m\phi\Big]+(B^{\mu}-\bar{B}^{\mu}+2\partial^{\mu}\phi)\partial_{\mu}\lambda,
 \end{eqnarray}
which clearly demonstrates that, on the r.h.s, we have the total spacetime derivative terms and the terms that contain the CF-type restrictions on our theory.
In exactly similar fashion, when we apply $s_{ab}$ on ${\cal L}_B$, we obtain the following:
\begin{eqnarray}
s_{ab}\mathcal{L}_{B}&=&\partial_{\mu}\Big[\rho \bar{B}^{ \mu} -(\partial^{\mu}\bar{C}^{\nu}-\partial^{\nu}\bar{C}^{\mu})B_{\nu}-2(\partial_{\nu}B^{\nu \mu})\rho -2m\rho \phi^{\mu}-(\partial^{\mu}\bar{C}-m\bar{C}^{\mu})B \nonumber\\ 
&-&\lambda \partial^{\mu}\bar{\beta} -m\varepsilon^{\mu \nu \sigma}\bar{C}_{\nu}\partial_{\sigma}\tilde{\varphi} \Big]+(\partial^{\mu}\bar{C}^{\nu}-\partial^{\nu}\bar{C}^{\mu})\partial_{\mu}\Big(B_{\nu}-\bar{B}_{\nu}+2\partial_{\nu}\phi\Big)\nonumber\\ 
&+&m(\partial^{\mu}\bar{C}-m\bar{C}^{\mu})(B_{\mu}-\bar{B}_{\mu}+2\partial_{\mu}\phi)+m\rho(B+\bar{B}-2m\phi) \nonumber\\
 &+&(\partial^{\mu}\bar{C}-m\bar{C}^{\mu})\partial_{\mu}\Big(B+\bar{B}-2m\phi\Big)+(B^{\mu}-\bar{B}^{\mu}+2\partial^{\mu}\phi)\partial_{\mu}\rho.
\end{eqnarray}
The r.h.s. of the above equation also shows that we have the total spacetime derivative terms plus terms that explicitly incorporate the CF-type
restrictions. Thus, it is crystal clear that, if we invoke the validity of the CF-type restrictions 
(i.e. $B + \bar B - 2 \,m \,\phi = 0, \; B_\mu- \bar B_\mu + 2 \,\partial_\mu\phi = 0$), we observe that ${\cal L}_B$ 
and  ${\cal L}_{\bar B}$ {\it both} respect {\it both} (i.e. BRST and anti-BRST) symmetry transformations on the submanifold of fields which
is defined by the the field equations corresponding to the CF-type restrictions.

We end this subsection with the following concluding remarks. First, we observe that, as far as the RRST and anti-BRST symmetries are concerned,
the Lagrangian densities ${\cal L}_B$ and  ${\cal L}_{\bar B}$ are {\it equivalent} provided we take into account the validity of 
the CF-type restrictions. Second,
these Lagrangian densities are {\it coupled} 
(cf. Subsec. 5.1) because they are connected to each-other by the CF-type restrictions (modulo a total spacetime derivative term).
Finally, we note that the existence of the non-trivial CF type restrictions {\it always} implies the existence of the coupled (but equivalent) Lagrangian densities
that respect both (i.e. BRST and anti-BRST) symmetries {\it together}.\\

\subsection{Absolute Anticommutatity Property: Significance}

For a given gauge theory with the local, continuous and infinitesimal gauge symmetry transformations, there exist two nilpotent symmetry transformations 
(within the framework of the BRST formalism) which are known as the BRST and anti-BRST symmetry transformations. The nilpotency property encodes the 
{\it fermionic} nature of these symmetries  which implies that, under these symmetry transformations, the fermionic fields 
of the BRST-quantized theory transform to the bosonic fields and vice-versa. However, these symmetries are {\it not} like the $\mathcal{N} = 2$
SUSY symmetry transformations because (i) the SUSY transformations do {\it not} absolutely anticommute with each-other, and (ii) the BRST and anti-BRST
symmetry transformations anticommute with each-other due to the existence of the (non-)trivial CF-type restrictions on the BRST quantized theory.
In fact, the absolute anticommutativity property of the (anti-)BRST symmetries is one of the sacrosanct properties of the BRST formalism.
Physically, this property encodes the linear independence of the BRST and anti-BRST symmetry transformations. For instance, in the case of 
our present {\it modified} 3D Abelian 2-form BRST-quantized theory, we observe the following:
\begin{eqnarray}
&& \Big\{s_{b},\;s_{ab}\Big\}B_{\mu \nu}=\partial_{\mu}(B_{\nu}-\bar{B}_{\nu})-\partial_{\nu}(B_{\mu}-\bar{B}_{\mu}),  \nonumber\\
&& \Big\{s_{b},\;s_{ab}\Big\}\phi_{\mu}=\partial_{\mu}(B+\bar{B})+ m\, (B_{\mu}-\bar{B}_{\mu}), \nonumber\\
&& \Big\{s_{b},\;s_{ab}\Big\}\Phi = 0,  \qquad
\Phi =\bar{C}_{\mu}, C_{\mu}, B_{\mu}, \bar{B}_{\mu}, B,\beta, \bar{\beta}, C, \bar{C}, \phi, \tilde{\varphi},  \bar{B}, {\cal B}.
\end{eqnarray}
The above equation demonstrates that the absolute anticommutativity property (i.e. $\{s_b, \, s_{ab} \} = 0$) of the BRST
and anti-BRST symmetry transformation operators [cf. Eqs. (26),(49),(69)]
is satisfied for all the fields of our theory provided we take into account the validity of the CF-type restrictions 
(i.e. $B + \bar B - 2 \,m\, \phi = 0, \; B_\mu- \bar B_\mu + 2 \,\partial_\mu\phi = 0$).

We end this subsection with the following comments. First of all, we note that the absolute anticommutativity property  (i.e. $\{s_b, \, s_{ab} \}  = 0$) is
automatically satisfied for the generic 
field $\Phi =\bar{C}_{\mu}, C_{\mu}, B_{\mu}, \bar{B}_{\mu}, \beta, \bar{\beta}, C, \bar{C}, \phi, \tilde{\varphi}, B, \bar{B}, {\cal B}$. However, for the 
gauge field and St${\ddot u}$ckelberg vector field, we have to invoke the validity of the CF-type restrictions.  This happens because the
gauge field $B_{\mu\nu}$ and the St${\ddot u}$ckelberg vector field $\phi_\mu$ are the most basic fields of our theory. Rest of the fields
appear in the theory because of the BRST quantization scheme of the St${\ddot u}$ckelberg-modified massive gauge theory. Second, the 
{\it (non-)trivial} CF-type restrictions are essential in the BRST-quantized theory because the independent identity of the BRST and anti-BRST
symmetries (and corresponding conserved charges) crucially depends on them. Furthermore, we have established (in our earlier work (see, e.g. [35]
and references therein)
that the CF-type restrictions are connected with the geometrical objects called gerbes. Finally, the independent nature of the BRST and
anti-BRST charges becomes  crystal  clear in the field-theoretic models for Hodge theory (see. e.g. [14,15]) where we observe 
that there is two-to-one mapping between the 
{\it six} conserved charges of the theory and {\it three} de Rham cohomological operators of differential geometry. We dwell a bit on this {\it mapping} aspect
in our next section where we have cited a {\it new} reference on a 3D field-theoretic example for Hodge theory. In this very recent work (where one of the
authors of our present manuscript is a co-author), 
such two-to-one mapping has been discussed in an elaborate and transparent manner.\\

%6
\section{Conclusions}

In our present endeavor, we have shown that under the local, continuous and infinitesimal (i) classical gauge symmetry transformations (4), and (ii) quantum (anti-)BRST symmetry transformations [cf. Eqs. (49),(26)], the kinetic terms of the 3D massive Abelian 2-form gauge field 
and the pseudo-scalar field ($\tilde{\varphi}$) remain invariant. The classical gauge symmetry transformations (4) are generated by a set of 
{\it four} first-class constraints [cf. Eq. (14)] that have been shown to exist on our theory. The (anti-)BRST invariance exists in our 
present theory because the gauge-fixing terms (for the massive Abelian 2-form field $B_{\mu \nu}$ as well as 
St${\ddot u}$ckelberg's vector field $\phi_{\mu}$) along with the FP-ghost terms transform in such a manner that the total (anti-)BRST invariant Lagrangian densities transform to the {\it total} spacetime\\
derivatives  [cf. Eqs. (50),(25)]. As a consequence, the action integrals, corresponding to the above (anti-)BRST invariant Lagrangian densities, respect the (anti-)BRST symmetry transformations. We have performed this exercise, for the first-time, in the case of an odd-dimensional (i.e. $D=3$) 
{\it modified} massive Abelian 2-form gauge theory.

One of the highlights of our present investigation is the observation that the pseudo-scalar field (i.e. $\tilde{\varphi}$) of our theory remains inert (i.e. $\delta_{g} \tilde{\varphi}=0, s_{(a) b} \tilde{\varphi}=0$) as far as the classical gauge symmetry transformations (4) and the (anti-)BRST symmetry transformations (49) and (26), respectively, are concerned. However, its presence in the 
{\it two} of the total four first-class constraints [cf. Eq. (14)] is an undeniable truth. According to Dirac's quantization conditions, the operator forms of these constraints must annihilate the physical states (in the total quantum Hilbert space of states) at the quantum level (see, e.g. [38] for details). We have been able to establish this fact, very briefly (cf. Appendix B below), where we have demanded that the physical states (in the total quantum Hilbert space of states) are 
{\it those} that are annihilated (i.e. $Q_{B}\, |phys >=0$) by the conserved and nilpotent version of the BRST charge $Q_{B}$. It is worthwhile to point out, in
the context of our present discussions, that the pseudo-scalar field (i.e. $\tilde{\varphi}$) does transform under the (anti-)co-BRST symmetry transformations which have been shown to exist, within the framework of BRST formalism, in the contexts of (i) the modified 2D massive Abelian 1-form 
(i.e. Proca) theory (see, e.g. [15]), and (ii) the modified massive 4D Abelian 2-form theory (see, e.g. [14,25]). These models, as mentioned earlier, are the field-theoretic examples for Hodge theory  which provide the physical realizations of the set of {\it three} de Rham cohomological operators of differential geometry at the {\it algebraic} level.

Against the backdrop of the discussions in the above paragraph, we would like to lay emphasis on the fact that the pseudo-scalar field (having the negative 
kinetic term and a well-defined mass) is very important field in our theory because (i) it turns out to be one of the possible candidates for 
dark matter (see, e.g. [33,34] and references therein),
and (ii) it plays a crucial role in the realm of cyclic, bouncing and self-accelerated cosmological models of the Universe 
(see, e.g. [27-32]). We speculate that
its {\it massless} limit (when {\it it} would
be endowed with {\it only} a negative kinetic term)  will be a possible candidate for dark energy because it would automatically lead to
the existence of negative pressure on the theory. The {\it latter} is one of the characteristic features of dark energy which is responsible for 
the modern observations of the accelerated expansion of the Universe (see, e.g. [49-51]). We further speculate that the pseudo-scalar field 
(having the negative kinetic term) will be 
one of the most fundamental fields that would correspond to the ``phantom''  and/or ``ghost'' fields of the cosmological models of the Universe
(see, e.g. [27-32] for details).

The existence of the (anti-)BRST invariant CF-type restriction(s) is the hallmark of a properly BRST-quantized theory. In this context, it is pertinent
to point out that the {\it trivial} CF-type restriction\footnote{The {\it trivial} CF-type condition, in the context of the BRST approach to any arbitrary
dimensional Abelian 1-form gauge theory, turns out to the limiting case of the celebrated CF-condition [36] that exists in the context of the BRST approach 
to any arbitrary dimensional  non-Abelian 1-form gauge theory which is responsible for the existence of the coupled (but equivalent)
(anti-)BRST invariant Lagrangian densities.}
%and (ii) the validity of the BRST and anti-BRST symmetry transformations [...].}
 exists in the simple cases of the BRST approach to the Abelian 1-form gauge theory, a system
of rigid rotor, etc., which are described by a {\it single} (anti-) BRST invariant Lagrangian density/Lagrangian  that respects the (anti-)BRST
symmetry transformations and their absolute anticommutativity property is satisfied automatically. As far as our 3D modified massive Abelian 2-form
theory is concerned, we have shown the existence of the non-trivial CF-type restrictions 
which are responsible for the existence of the coupled (but equivalent)
BRST and anti-BRST invariant Lagrangian densities. Furthermore, in the proof of the absolute anticommutativity property of the BRST and
anti-BRST transformations, we have been compelled to invoke the sanctity of the non-trivial CF-type resections. All these aspects 
have been discussed in our Sec. 5 (and we have been able to derive the CF-type restrictions from different theoretical angles).

One of the very promising future endeavor for us would be look into the possibility of the existence of the nilpotent (i.e. $s_{(a) d}^{2}=0$) (anti-)co-BRST/(anti-)dual-BRST symmetry transformations ($s_{(a) d}$) for our 3D modified massive Abelian 2-form theory. It is expected that, under 
{\it these} symmetry transformations, the total gauge-fixing terms for the massive gauge field as well as St${\ddot u}$ckelberg's vector field would remain invariant. In other words, we shall have the following kinds of (anti-)dual-BRST transformations (see, e.g. [52-54]):
\begin{eqnarray}
&& s_{a d} B_{\mu \nu}=-\varepsilon_{\mu \nu \sigma}\, \partial^{\sigma} C, \qquad s_{a d} C = 0, \qquad 
s_{a d} \,\phi_{\mu}= \varepsilon_{\mu\nu\sigma}\, \partial^\nu C^\sigma, \nonumber\\ 
&& s_{ad} C_\mu = \partial_\mu \beta, \qquad s_{a d} \,\phi=0, \qquad s_{a d}\left(\partial^{\nu} B_{\nu \mu}\right)=0, \nonumber\\
&& s_{d} B_{ \mu \nu}=-\varepsilon_{\mu \nu \sigma} \,\partial^{\sigma} \bar{C}, \qquad s_{d} \bar{C}=0, \qquad s_{d} \,\phi_{\mu}= \varepsilon_{\mu\nu\sigma}\, \partial^\nu \bar C^\sigma, \nonumber\\
&& s_d \bar C_\mu = \partial_\mu \bar \beta, \qquad s_{d} \,\phi=0, \qquad s_{d}\left(\partial^{\nu} B_{\nu \mu}\right)=0.
\end{eqnarray}
On top of these symmetry transformations, we have to find out the appropriate transformations for the (anti-)ghost fields as well as for the pseudo-scalar field so that the total transformations on (i) the kinetic terms for the gauge field as well as the pseudo-scalar field, and (ii) the ghost-sector of the total (anti-)BRST invariant Lagrangian densities sum-up to the total spacetime derivatives. Thus, in our future endeavor, we wish to prove that our present odd dimensional 
(i.e. $D=3$) modified massive Abelian 2 -form theory is {\it also} a field-theoretic model for Hodge theory. It is gratifying to state
that a {\it limiting} case of our present endeavor (which turns out to be a {\it combined} system of the free Abelian 1-form and 2-form gauge theories) has
already been proven [52] to be an example for Hodge theory and its constraint structures and elaborate BRST analysis have been performs in [53].  
Another direction that can be followed by us is the constraint analysis of this theory, in an elaborate manner, within the framework of BRST formalism. We have, very concisely, discussed a bit of this aspect of our theory in our Appendix B to show that the existence of the pseudo-scalar field (i.e. $\tilde \varphi $)
in our theory is important because it is present in the operator form of the {\it two} constraints which annihilate the physical states (i.e. $ |phys>$) of our theory at the quantum level (cf. Appendix B). In a very recent work [54], we have been able to prove a 3D field-theoretic model to be 
an example for Hodge theory\footnote{We have {\it six} continuous symmetries in {\it this} theory out of which {\it four} are
fermonic [i.e. (anti-)BRST and (anti-)co-BRST] in nature and two are bosonic (i.e. a {\it unique} bosonic and ghost-scale). These symmetries lead to
the derivations of the appropriate conserved charges which are the generators for the above {\it six} continuous symmetry transformations. The key
observations connected with these charges are (i) the BRST and anti-co-BRST charges raise the ghost number 
of a given quantum state by {\it one}, (ii) the anti-BRST and co-BRST charges
lower the ghost number of a given state by {\it one}, and (iii) the {\it unique} bosonic charge keeps the ghost number {\it intact} for a given state.
These observations are just like the operations of the exterior derivative, co-exterior derivative and Laplacian operator on a given differential
form of a given degree (see, e.g. [21-14]). Hence, there is two-to-one mappings between the six conserved charges and three
cohomological operators in the case of an example for Hodge theory (see. e.g. [54] for details). }
where the transformations of the type (74) have been taken into account where the (anti-)co-BRST symmetry transformations for the pseudo-scalar field have
been written.
%, and (ii) the nilpotent version of the BRST charge is superior to the Noether non-nilpotent BRST charge as far as the
%constraint analysis is concerned ( 
%Yet another direction of our interest is to apply the BRST formalism to the 4D modified massive Abelian 3-form theory for which the 
%properly gauge-fixed Lagrangian density has been written down in our earlier work [29]. 
We are currently
very much involved with  the above cited  problems and our results will be reported elsewhere.

At this juncture, it is worthwhile to mention
that the theoretical material, presented in our present endeavor, has already been published as a preprint in [55].\\

%\vskip 0.5cm

\noindent
{\bf Acknowledgments}\\

\noindent
Our present work was initiated  when one of us (RPM) was invited to visit the School of Physics, University of Hyderabad (UoH). 
He is immensely grateful to the  IoE-UoH- IPDF (EH) scheme of UoH for his travel support and warm local hospitality. 
All the authors gratefully acknowledge a crucial technical help, provided by A. K. Rao, 
in the preparation  of this manuscript.\\

\vskip 0.1cm

\noindent
{\bf  Data Availability Statement}\\

\vskip 0.1cm

\noindent
Data sharing not applicable to this article as no datasets were generated or analysed during the current study.\\

\vskip 0.1cm
\noindent
{\bf Conflicts of Interest} \\

\vskip 0.1cm

\noindent
The authors declare that there are no conflicts of interest. \\

%\vskip 0.1cm

\noindent
{\bf Funding Statement}} \\

%\vskip 0.1cm

\noindent
No funding was received for this research.\\

\noindent
{\bf Authors Contributions} \\

S. K. Panja: Writing-review and editing, Methodology.\\

E. Harikumar: Conceptualization.\\

R. P. Malik: Investigation. \\

%\vskip 0.1cm

\noindent
{\bf Authors Confirmation}\\

\noindent
All the authors agree to be accountable for the research presented.\\

%\vskip 0.1cm

\noindent
{\bf Declaration}\\

\noindent
All the authors declare that this manuscript is available as a preprint on the arXiv whose link is: https://arxiv.org/pdf/2405.15588.\\

\vskip 0.3cm

\begin{center}
{\bf Appendix A: \bf  On the Standard BRST Algebra}\\
\end{center}

\vskip 0.5cm

\noindent
In addition to the infinitesimal, continuous and off-shell nilpotent (i.e. fermionic) BRST and anti-BRST symmetry transformations that have been discussed in Secs. 3 and 4, respectively, we {\it also}  have 
the following  set of {\it bosonic} ghost-scale transformations, namely;
\[
C_{\mu} \rightarrow e^{+\Sigma} C_{\mu}, \quad \bar{C}_{\mu} \rightarrow e^{-\Sigma} \bar{C}_{\mu}, \quad \beta \rightarrow e^{+2 \Sigma} \beta, \quad \bar{\beta} \rightarrow e^{-2 \Sigma} \bar{\beta}, 
\]
\[
C \rightarrow e^{+\Sigma} C, \quad \bar{C} \rightarrow e^{-\Sigma} \bar{C}, \quad \lambda \rightarrow e^{+\Sigma} \lambda, \quad \rho \rightarrow e^{-\Sigma} \rho,
\]
\[ 
\Phi \rightarrow e^{0} \Phi \quad\left(\Phi=B_{\mu \nu}, B_{\mu}, \mathcal{B}, B, \bar{B}_{\mu}, \bar{B}, \phi_{\mu}, \phi, \tilde{\varphi}\right),
\eqno(A.1)
\]
where $\Sigma$ is a spcetime-independent (i.e. global) scale transformation parameter and the numerals in the exponents correspond to the ghost numbers for the fields. It is clear that all the fields of the non-ghost sector (i.e. $B_{\mu \nu}, B_{\mu}, \mathcal{B}, B, \bar{B}_{\mu}, \bar{B}, \phi_{\mu}, \phi, \tilde{\varphi}$) carry the ghost number equal to zero. Hence, in the exponent, corresponding to the ghost-scale symmetry transformation on the generic field $\Phi$, we have {\it zero} as the numeral. For the sake of brevity, we take into account $\Sigma=1$ so that the infinitesimal version $\left(s_{g}\right)$ of the above ghost-scale symmetry transformations are as follows
\[
s_{g} C_{\mu}=+C_{\mu}, \quad s_{g} \bar{C}_{\mu}=-\bar{C}_{\mu}, \quad s_{g} \beta=+2 \beta, \quad s_{g} \bar{\beta}=-2 \bar{\beta}, \]\[
s_{g} C=+C, \quad s_{g} \bar{C}=-\bar{C}, \quad s_{g} \lambda=+\lambda, \quad s_{g} \rho=-\rho, \quad s_{g} \Phi=0,
\eqno(A.2)
\]
where it can be readily checked that $s_{g}$ is bosonic (i.e. $s_{g}^{2} \neq 0$) in nature. Under these infinitesimal transformations, the (anti-)BRST invariant Lagrangian densities remain invariant (i.e. $s_{g} \mathcal{L}_{B}=0, s_{g} \mathcal{L}_{\bar{B}}=0$). As a consequence, according to Noether's theorem, we have the following expression for the ghost current $J_{(g)}^{\mu}$, namely;
\[
J_{(g)}^{\mu}= 2\, \beta \,\partial^{\mu} \bar{\beta}- 2\, \bar{\beta} \,\partial^{\mu} \beta
+ \left(\partial^{\mu} C^{\nu}-\partial^{\nu} C^{\mu}\right) \bar{C}_{\nu} + \lambda\, \bar{C}^{\mu} 
+ \left(\partial^{\mu} \bar{C}^{\nu}-\partial^{\nu} \bar{C}^{\mu}\right) C_{\nu}
- \rho \, C^{\mu}\]
\[
-\left(\partial^{\mu} \bar{C}- m \bar{C}^{\mu}\right) C-\left(\partial^{\mu} C- m  C^{\mu}\right) \bar{C}.
\eqno(A.3)
\]
The conservation law ($\partial_{\mu} J_{(g)}^{\mu}=0$) can be proven readily by using the EL-EoMs (28) that have been derived from the ghost-sector of the (anti-)BRST invariant Lagrangian densities. The expression for the conserved ghost charge $Q_{g}=\int d^{2} x J_{(g)}^{0}$ is as follows:
\[
Q_{g}=\int d^{2} x\left[2 \beta \dot{\bar{\beta}}-2 \bar{\beta} \dot{\beta}+\left(\partial^{0} C^{i}-\partial^{i} C^{0}\right) \bar{C}_{i}
+ \lambda \,\bar{C}^{0} - \rho\, C^{0} + \left(\partial^{0} \bar{C}^{i}-\partial^{i} \bar{C}^{0}\right) C_{i}\right. 
\]
\[
\left.-\left(\partial^{0} \bar{C}-m \bar{C}^{0}\right) C-\left(\partial^{0} C-m C^{0}\right) \bar{C}\right]. 
\eqno(A.4)
\]
The above charge is the generator for the infinitesimal ghost transformations (A.2) if we express the ghost charge in terms of the 
 canonical conjugate  momenta w.r.t. the basic (anti-)ghost fields of our theory and use the general equation (31) with the replacements: $s_{b} \rightarrow s_{g}, Q_{b} \rightarrow Q_{g}$ and use only the commutator (i.e. $[\,,\,]_{(-)}$) on the r.h.s. of (31).

Taking into account the property of the ghost charge $Q_{g}$ as the generator for the infinitesimal ghost-scale transformations $s_{g}$, we further note that the following relationships between the ghost charge and the non-nilpotent version (i.e. $Q_{(a) b}^{2} \neq 0$)
and nilpotent  version (i.e. $ Q_{(A) B}^{2}=0$) of the (anti-)BRST charges (i.e. $Q_{(a) b}, Q_{(A) B}$) are true, namely;
\[
s_{g} Q_{b}=-i\left[Q_{b}, Q_{g}\right]=+Q_{b}  \Rightarrow  i\left[Q_{g}, Q_{b}\right]=+Q_{b}, 
\]
\[
s_{g} Q_{a b}=-i\left[Q_{a b}, Q_{g}\right]=-Q_{a b}  \Rightarrow  i\left[Q_{g}, Q_{a b}\right]=-Q_{a b},
 \]
 \[
s_{g} Q_{B}=-i\left[Q_{B}, Q_{g}\right]= + \,Q_{B}  \Rightarrow  i\left[Q_{g}, Q_{B}\right]=+Q_{B},
\]
\[
s_{g} Q_{A B}=-i\left[Q_{A B}, Q_{g}\right]= -\, Q_{A B}  \Rightarrow  i\left[Q_{g}, Q_{A B}\right]=-Q_{A B},
\eqno(A.5)
\]
which demonstrate that the Noether non-nilpotent (anti-)BRST charges $Q_{(a) b}$ and the nilpotent versions of the (anti-)BRST charges $Q_{(A) B}$ obey the same kinds of algebras with the conserved ghost charge $Q_{g}$. However, we know that the off-shell nilpotency property of the (anti-)BRST charges is very important from the point of view of (i) the BRST cohomology, and (ii) the physicality criteria (and their consistency with the Dirac quantization conditions for the systems with constraints). Thus, the standard BRST algebra is obeyed amongst the nilpotent versions (i.e. $Q_{(A) B}^{2}=0$) of the (anti-)BRST charges $Q_{(A) B}$ and the conserved ghost charge $Q_{g}$. This well-known and beautiful algebra is as follows:
\[
Q_{B}^{2}=0, \quad Q_{A B}^{2}=0, \quad i\left[Q_{g}, Q_{B}\right]= +\, Q_{B}, \quad i\left[Q_{g}, Q_{A B}\right]= - \,Q_{A B}.
\eqno(A.6)
\]
The above algebra encodes the fact that the ghost numbers of the (anti-)BRST charges are $(-1)+1$, respectively. In other words, the BRST transformation increases the ghost number of a field by one [cf. Eq. (26)]. On the other hand, the ghost number decreases by one  for a field on which the anti-BRST transformation operates [cf. Eq. (49)].\\

\vskip0.5cm

\begin{center}
{\bf Appendix B: \bf  On the Physicality Criterion w.r.t. $Q_{B}$}\\
\end{center}

\vskip 0.5cm

\noindent
The purpose of our present Appendix is to discuss briefly
the consequences of the physicality criterion w.r.t. the nilpotent version of the BRST charge $Q_{B}$ and to establish its superiority over the non-nilpotent version of the Noether BRST charge $Q_{b}$. To be precise, we show that all the primary as well as the secondary constraints of our theory annihilate the physical states when we demand that the physical states (i.e. $|phys >$) are those (in the total quantum Hilbert space of states) that are annihilated (i.e. $Q_{B}\, |phys >=0$) by the conserved and {\it nilpotent} version of the BRST charge $Q_{B}$
operator. Before we shall do this exercise, we would like to point out that there are no explicit primary constraints for the BRST invariant Lagrangian density (23). Rather, they have been traded with the Nakanishi-Lautrup type auxiliary fields because we observe the following:
\[
\Pi_{(B)}^{\mu \nu}=\frac{\partial \mathcal{L}_{B}}{\partial\left(\partial_{0} B_{\mu \nu}\right)}=\frac{1}{2} \varepsilon^{0 \mu \nu} \mathcal{B}+\frac{1}{2}\left(\eta^{0 \mu} B^{\nu}-\eta^{0 \nu} B^{\mu}\right), 
\]
\[
\Pi_{(\phi)}^{\mu}=\frac{\partial \mathcal{L}_{B}}{\partial\left(\partial_{0} \phi_{\mu}\right)}=m B^{0 \mu}-\Sigma^{0 \mu}-\eta^{0 \mu} B. \eqno(B.1)
\]
The expressions in (B.1) demonstrate that we have the following:
\[
\Pi_{(B)}^{0 i}=\frac{1}{2} B^{i}, \qquad \Pi_{(B)}^{i j}=\frac{1}{2} \varepsilon^{0 i j} \mathcal{B}, \qquad \Pi_{(\phi)}^{0}=-B, 
\qquad \Pi_{(\phi)}^{i}=m B^{0 i}-\Sigma^{0 i}. 
\eqno(B.2)
\]
These observations corroborate our statement that the primary constraints $\Pi_{(B)}^{0 i} \approx 0$ 
and $\Pi_{(\phi)}^{0} \approx 0$ [cf. Eq. (9)]  of our starting 
Lagrangian density $\mathcal{L}_{(0)}^{(M S)}$ have been traded with the Nakanishi-Lautrup auxiliary fields $B^{i}$ and $B$, respectively [cf. Eq. (B.2)].

A close and careful look at the (anti-)BRST invariant Lagrangian densities reveals that the (anti-)ghost fields are decoupled, right from the beginning, from the {\it rest}
 of the theory as there is no interaction between the bosonic  $\left(B_{\mu \nu}, \phi_{\mu}, \phi, \tilde{\varphi}\right)$ 
fields (with the ghost number zero) and the (anti-)ghost fields 
(which carry the non-zero ghost numbers) of our theory. Hence the quantum Hilbert space of states of our theory is a direct product (see, e.g. [56])
of the physical states (i.e. $|phys >$) and the ghost states (i.e. $|ghost >$. When we demand 
the physicality criterion on the quantum Hilbert space of states w.r.t. the nilpotent version of the BRST charge [cf. Eq. (46)], the ghost fields act only on the ghost states (i.e. $|ghost >$) and produce the non-zero results. Hence, to satisfy the condition $Q_{B}\, | phys >=0$, all the fields and/or the specific combination of them (with the zero ghost number) that are (i) present in the expression for $Q_{B}$, and 
(ii) associated with the {\it basic} (anti-)ghost fields, must annihilate the physical states (i.e. $|phys >$). As a consequence, the following operator conditions emerge out on the physical states from the requirement: $Q_{B}\, | phys >=0$ w.r.t. the nilpotent version of the BRST charge, namely;
\[
B \,|phys> = 0, \qquad B^{i} \,|phys> = 0, \qquad \left(\partial^{0} B + m\, B^{0}\right) \,|phys> = 0, 
\]
\[
\left(\partial^{0} B^{i}-\partial^{i} B^{0}\right) \,|phys> = 0.
\eqno(B.3)
\]
It is pretty obvious, from (B.2), that $B \,|phys> = 0,\, B_{i} \,|phys> = 0$ imply that the operator forms of the primary constraints annihilate (i.e. $\Pi_{(\phi)}^{0} \,|phys> = 0, \, \Pi_{(B)}^{0 i} \,|phys> = 0$) the physical states of our theory. Let us now concentrate on the last two entries of (B.3) and\\
prove that they imply that the secondary constraints of our theory annihilate the physical states. For this purpose, we have to take into account the following EL-EoMs that emerge out from the non-ghost sector of the BRST-invariant Lagrangian density, namely;
\[
\partial_{\mu}\left(m B^{\mu \nu}\right.  \left.-\Sigma^{\mu \nu}\right)=m B^{\nu}+\partial^{\nu} B, 
\]
\[
\frac{1}{2} \varepsilon^{\mu \nu \sigma} \partial_{\mu} \mathcal{B}+\frac{1}{2}\left(\partial^{\nu} B^{\sigma}-\partial^{\sigma} B^{\nu}\right)  =-\frac{m}{2}\left(m B^{\nu \sigma}-\Sigma^{\nu \sigma}\right)+\frac{m}{2} \varepsilon^{\nu \sigma \eta} \partial_{\eta} \tilde{\varphi}.
\eqno(B.4)
\]
Taking the choice $\nu=0$ in the top entry of the above equation and the choosing $\nu=0, \sigma=i$ in the bottom entry, we obtain the following very useful relationships:
\[
\left(\partial^{0} B+m B^{0}\right)  =-\partial_{i}\left(m B^{0 i}-\Sigma^{0 i}\right) \equiv-\partial_{i} \Pi_{(\phi)}^{i}, 
\]
\[
\left(\partial^{0} B^{i}-\partial^{i} B^{0}\right)  =2\left(\partial_{j} \Pi_{(\phi)}^{j i}-\frac{m}{2} \Pi_{(\phi)}^{i}+\frac{m}{2} \varepsilon^{0 i j} \partial_{j} \tilde{\varphi}\right) 
\]
\[
 \equiv-2\left(\frac{m}{2} \Pi_{(\phi)}^{i}-\partial_{j} \Pi_{(\phi)}^{j i}-\frac{m}{2} \varepsilon^{0 i j} \partial_{j} \tilde{\varphi}\right).
  \eqno(B.5)
\]
The above beautiful relationships establish that the third and fourth conditions on the physical states (i.e. $|phys>$) in (B.3) are nothing but the conditions that the operator forms of the secondary constraints [cf. Eqs. (10),(12)] of our classical theory annihilate the physical states that exist in the total quantum Hilbert space of states.

We end this Appendix with the following closing remarks. First of all, we note that the physicality criterion [cf. Eq. (B.3)] and ensuing conditions on the physical states w.r.t. the {\it nilpotent} version of the BRST charge are consistent with the Dirac quantization conditions (for the quantization of systems that are endowed with constraints). Second, we have not taken into account the condition: $B^{0} \, |phys> $ from the physicality criteria w.r.t.the BRST charge $Q_{B}$ because the auxiliary field component $B^{0}$ is associated with the auxiliary ghost field $\lambda$ which is 
{\it not} a basic ghost field. Moreover, it (i.e. $B^{0}$) is the conjugate momentum corresponding to the scalar field $\phi$ which is non-zero and, therefore, it is not a constraint on our theory. Third, even though the pseudo-scalar field does not transform under the classical gauge transformations [cf. Eq. (4)] and nilpotent
(anti-)BRST transformations [cf. Eqs. (49),(26)] at the quantum level, its presence and importance in the constraints at the classical level
[cf. Eq. (14)] as well as at the quantum level [cf. Eqs. (B.5),(B.3)] can {\it not} be denied.
Fourth, it can be explicitly checked that the physicality criteria (i.e. $Q_{b}\,|phys >=0$) w.r.t. the non-nilpotent (i.e. $Q_{b}^{2} \neq 0$) Noether conserved charge $Q_{b}$ [cf. Eq. (30)] leads to the annihilation of the physical states {\it only} by the operator forms of the {\it primary}
 constraints [cf. Eq. (9)] of our classical theory. However, it does not lead to the annihilation of the physical states by the operator forms of the secondary constraints [cf. Eqs. (10),(12)] that are present on our classical theory. Finally, we would like to mention, in passing, that one can take into account the nilpotent version 
of the anti-BRST charge $Q_{AB}$ in the discussion of the physicality criterion. However, the results will be same as what we have obtained 
in the case of our present discussions w.r.t. the nilpotent version of the  BRST charge $Q_{B}$ (modulo some signs and symbols). \\

\vskip0.7cm

\begin{center}
{\bf Appendix C: \bf  On the BRST Quantization of Our 3D System}\\
\end{center}
 
\vskip0.7cm

\noindent
The central objective  of this Appendix is to discuss the BRST-quantization of our 3D field-theoretic model of the St{\" u}ckelberg-modified massive 
Abelian 2-form theory where the basic equal-time canonical commutators (in the natural units: $\hbar = c = 1 $) have already been mentioned in equation (17).
These {\it quantum} level non-zero commutators correspond to the {\it non-ghost} sector of the BRST-invariant Lagrangian density ${\cal L}_{B}$ (cf. main body
of our text). It is worthwhile to mention, in this context, that {\it all} the canonical conjugate momenta w.r.t. the $B_{\mu\nu}$ and $\phi_\mu$ fields
are non-zero [cf. Eq. (B.2)] in our Appendix B. The other non-zero momentum $\Pi_{(\tilde \varphi)} $ 
[cf. Eq. (8)] can be readily computed from the gauge-fixed
Lagrangian density (23). Thus, we find that there is {\it no} problem in the covariant canonical quantization (i.e. BRST-quantization) of our theory as far as
its {\it non-ghost} sector of the BRST-invariant Lagrangian density ${\cal L}_B$ (cf. the main  body of our text) is concerned.

At this juncture, to complete our whole discussion on the BRST-quantization scheme, we focus on the {\it ghost-sector} of our theory which is 
described by the Lagrangian density ${\cal L}_{(FP)}$ [cf. Eq. (24)]. In this context, it can be noted that we have the following canonical conjugate momenta
w.r.t. the fields $\beta, \bar \beta, C, \bar C, C_\mu, \bar C_\mu$ of the Lagrangian density ${\cal L}_{(FP)}$:
%\begin{eqnarray*}
\[
\Pi_{(\beta)}  =\frac{\partial \mathcal{L}_{(FP)}}{\partial\left(\partial_{0} \beta\right)}= \dot {\bar \beta}, \qquad 
\Pi_{(\bar \beta)}  =\frac{\partial \mathcal{L}_{(FP)}}{\partial\left(\partial_{0} \bar \beta\right)}= \dot { \beta}, \qquad
\Pi_{(C)}  =\frac{\partial \mathcal{L}_{(FP)}}{\partial\left(\partial_{0} C\right)}= +\, \big(\partial^0 {\bar C} - m \, \bar C^0 \big ), 
\]
\[ 
\Pi_{(\bar C)}  =\frac{\partial \mathcal{L}_{(FP)}}{\partial\left(\partial_{0} \bar C\right)}= -\, \big(\partial^0 {C} - m \,  C^0 \big ), 
\quad \Pi^\mu_{(C)}  =\frac{\partial \mathcal{L}_{(FP)}}{\partial\left(\partial_{0} C_\mu\right)}= -\, 
\big(\partial^0 {\bar C^\mu} - \partial^\mu \bar C^0 \big) + \eta^{\mu 0}\, \rho, 
\]
\[
\Pi^\mu_{(\bar C)}  =\frac{\partial \mathcal{L}_{(FP)}}{\partial\left(\partial_{0} \bar C_\mu\right)}= +\, 
\big(\partial^0 { C^\mu} - \partial^\mu  C^0 \big) + \eta^{\mu0}\, \lambda. \eqno(C.1)
\] 
%$~~~~~~~~~~~~~~~ (C.1)$ 
%\end{eqnarray*} 
It should be noted that, despite the subscripts being the same, the superscript(s) differentiate the conjugate moments w.r.t. the (anti-)ghost fields
$(\bar C)C$ and $(\bar C_\mu)C_\mu$ in the above expressions. In view of the equation (C.1), it is straightforward to obtain the following 
non-zero (anti)commutators (in the natural units: $\hbar = c = 1 $)
for the ghost-sector, namely;
%\begin{eqnarray*}
%$$
%\begin{aligned}
\[
{\left\{C (\vec{x}, t), \;\Pi_{(C)}(\vec{y}, t)\right\} }  = i\, \delta^{(2)}(\vec{x}-\vec{y}),\qquad
{\left\{\bar C (\vec{x}, t), \;\Pi_{(\bar C)}(\vec{y}, t)\right\} }  = i\, \delta^{(2)}(\vec{x}-\vec{y}), 
\]
%\end{aligned}
%$$
\[
{\left\{C_\mu (\vec{x}, t), \;\Pi^\nu_{(C)}(\vec{y}, t)\right\} }  = i\, \delta^\nu_\mu\, \delta^{(2)}(\vec{x}-\vec{y}),
\qquad
{\left\{\bar C_\mu (\vec{x}, t), \;\Pi^\nu_{(\bar C)}(\vec{y}, t)\right\} }  = i\, \delta^\nu_\mu\, \delta^{(2)}(\vec{x}-\vec{y}), 
\]
\[
{\left[\beta (\vec{x}, t), \;\Pi_{(\beta)}(\vec{y}, t)\right] }  = i \,\delta^{(2)}(\vec{x}-\vec{y}),
\qquad
{\left[\bar \beta (\vec{x}, t), \;\Pi_{(\bar \beta)}(\vec{y}, t)\right] }  = i \,\delta^{(2)}(\vec{x}-\vec{y}), \eqno(C.2)
\]
%$~~~~~~~~~~~~~~~~~~~~~~~~~~~~~~~~~~~~~~~~~~~~~~~~~~~~~~~~~(C.2)$
%\end{eqnarray*}
and the rest of the (anti)commutators are zero as per the canonical quantization procedure.

We end this Appendix with a few concluding remarks. First of all, we  defined the conjugate momenta in equations (8) and (B.2)
(cf Appendix B) for the perfectly BRST-invariant Lagrangian density ${\cal L}_B$ from its non-ghost sector [cf. Eq. (23)]. However, similar kinds of
conjugate momenta can be defined for the Lagrangian density (48)  which is a part of the perfectly anti-BRST invariant 
Lagrangian density ${\cal L}_{\bar B}$ and the non-zero commutators (17) can be derived for this Lagrangian density, too. The 
expressions for these conjugate momenta, of course,  will be {\it different}. Second, the FP-ghost part of the Lagrangian density [cf. Eq. (24)]
is the {\it same} for the {\it coupled} Lagrangian densities  ${\cal L}_B$ and ${\cal L}_{\bar B}$. Hence, the canonical 
(anti)commutators (C.2) will remain the same for both sets of the above {\it coupled} Lagrangian densities. Finally, the BRST-quantization scheme
is also known as the covariant canonical quantization scheme because of the fact that {\it none} of the components of the conjugate momenta
w.r.t. the gauge field $B_{\mu\nu}$ and the St{\" u}ckelberg field $\phi_\mu$ are {\it strongly} equal to zero within the framework
of BRST formalism [cf. Eq. (B.2)].  Hence, basic non-zero canonical (anti)commutators are well-defined in {\it this} quantization scheme.\\

\end{document}